\documentclass[useAMS,usenatbib]{mn2e}
\usepackage{amsmath}
\usepackage{amssymb}
\usepackage{threeparttable}
\usepackage{cases}
\usepackage{color}
\usepackage{graphicx}
\usepackage[hyperfootnotes=false]{hyperref}
\usepackage[figure]{hypcap}

\hypersetup{colorlinks, citecolor=blue, pdfduplex=DuplexFlipLongEdge}
\hypersetup{pdftitle=Light from the Inner Plunging Region of Accretion Discs, pdfauthor=Yucong Zhu, pdfsubject=Astrophysics}

\defcitealias{kulkarni11}{K2011}
\defcitealias{NT}{NT}

\title[Light from the Inner Plunging Region of Accretion Discs]
  {The Eye of the Storm:\\ Light from the Inner Plunging Region of Black Hole Accretion Discs}
\author[Y. Zhu et al.]
  {Yucong Zhu$^1$\thanks{E-mail:\newline \hbox{yzhu@cfa.harvard.edu (YZ);} \hbox{swd@cita.utoronto.ca (SWD);} \hbox{rnarayan@cfa.harvard.edu~(RN);} \hbox{akulkarni@cfa.harvard.edu~(AKK);} \hbox{rpenna@cfa.harvard.edu~(RFP);} \hbox{jem@cfa.harvard.edu~(JEM)}},
 Shane W. Davis$^2$\footnotemark[1], Ramesh Narayan$^1$\footnotemark[1],
\newauthor
  Akshay K. Kulkarni$^1$\footnotemark[1], Robert F. Penna$^1$\footnotemark[1], Jeffrey E. McClintock$^1$\footnotemark[1]\\
  $^1$Harvard-Smithsonian Center for Astrophysics, 60 Garden Street, Cambridge, MA 02138, USA\\
  $^2$Canadian Institute for Theoretical Astrophysics. Toronto, ON M5S3H4, Canada}
\date{\today}

\begin{document}

\maketitle

\label{firstpage}


\begin{abstract}

It is generally thought that the light coming from the inner plunging region of black hole accretion discs contributes negligibly to the disc's overall spectrum, i.e. the plunging fluid is swallowed by the black hole before it has time to radiate.  In the standard disc model used to fit X-ray observations of accretion discs, the plunging region is assumed to be perfectly dark.  However, numerical simulations that include the full physics of the magnetized flow predict that a small fraction of the disc's total luminosity emanates from the plunging region.  We investigate the observational consequences of this neglected inner light.  We compute radiative transfer based disc spectra that correspond to 3D general relativistic magnetohydrodynamic simulated discs (which produce light inside their plunging regions).  In the context of black hole spin estimation, we find that the neglected inner light only has a modest effect (this bias is less than typical observational systematic errors).  For rapidly spinning black holes, we find that the combined emission from the plunging region produces a weak power-law tail at high energies.  This indicates that infalling matter is the origin for some of the `coronal' emission observed in the thermal dominant and steep power-law states of X-ray binaries.

\end{abstract}
\begin{keywords}
accretion, accretion discs -- black hole physics -- MHD --   methods: numerical -- radiative transfer -- X-rays: binaries
\end{keywords}

\section{Introduction}\label{sec:intro}

The ubiquity and simplicity of black holes in our universe make them truly marvelous objects of study.  The complete physics of each and every black hole (BH) can be distilled down to just two numbers\footnote{In principle, BH charge is an independent parameter; however, we do not expect astrophysical black holes to retain any significant charge.}: the black hole's mass $M$ and angular momentum $J$, the latter of which is usually expressed as a dimensionless spin parameter $a_*=J/(GM^2/c)$.  This implies that all physical theories that involve black holes, e.g. the link between BH spin and jet power \citep{blandford77}, gamma ray bursts and spinning BHs \citep{macfadyen99}, models of quasi-periodic oscillations \citep{abram01}, must connect in some way to these two numbers.     

For binary systems, the task of measuring mass is relatively straightforward; so long as one can obtain the period, orbital velocity, orbital geometry (i.e. inclination and eccentricity), and mass for a companion orbiting a black hole, one can immediately obtain the black hole mass using only Newtonian gravity \citep[for a recent example, see][]{orosz11b}.  The game of measuring black hole masses has been played as early as 1972 for Cygnus X-1 \citep{webster72,bolton72}, and to date we have robust mass estimates for about 20 stellar mass binary black hole systems \citep{remillard06,orosz07,orosz09,orosz11a,orosz11b,cantrell10}, and 64 supermassive black holes \citep{gultekin09,graham11}.

Despite this success in measuring black hole mass, spin has been a more difficult quantity to obtain.  The spin of an object only makes itself known at very short distances through a general relativistic effect known as frame dragging. Thus to probe black hole spin, we must rely on observations of matter that is very close to the BH horizon.  In practice, the only available probe is the accretion disc orbiting the black hole, which transitions from nearly circular orbits to plunging at a special location known as the innermost stable circular orbit (ISCO).  By observing the light given off by the accretion disc,  it is possible to determine this transition radius.  When the ISCO is expressed in terms of the gravitational radius $r_g=GM/c^2$, it has a well-defined monotonic dependence on black hole spin (e.g. \citealt{shapiro83}). Therefore, a measurement of the ISCO size yields the BH spin if the mass is independently known.

For a black hole binary system, we measure the size of the ISCO by modelling the light given off by the accretion disc.  Objects inside the ISCO (hereafter referred to as the `plunging zone') cannot remain in stable circular orbits, and are forced to plunge into the black hole in just a few free-fall times.  Since the plunging time-scale is short, it is thought that fluid inside the plunging zone does not have time to radiate \citep{pagethorne74}, which means that if we could resolve an image of the accretion disc, then we would see a dark void corresponding to the plunging region in the centre.  In practice, we cannot resolve images\footnote{Although it has been recently proposed to use radio VLBI observations to test GR by resolving the apparent shape of SgrA$^*$ \citep{doeleman09}.} of accretion discs around black holes, and thus our only information comes in the form of X-ray spectra.

One method of estimating BH spin\footnote{Another commonly applied method is known as the `iron line' method, which estimates BH spin through spectral modelling of the Fe K$\alpha$ fluorescence line (\citealt{fabian89}, see \citealt{miller07} for a recent review).} works by fitting the spectral shape of the thermal continuum emission from the accretion disc to thereby estimate the radius of the ISCO (this method was pioneered by \citealt{zhang97a}; see Table \ref{tab:literatureSpinFits} for some recent results).  

This `continuum fitting' method uses the colour temperature of the thermal component, the distance to the source, and the received X-ray flux to obtain a characteristic emitting area for the disc.  Given a model for the radial dependence of the disc emission, this characteristic emitting area determines the ISCO radius.  The main drawback of the continuum fitting method is that we first need accurate estimates of the BH distance (to turn fluxes into areas), disc inclination (to turn the projected area into an ISCO radius), and BH mass (to get spin from the monotonic $r_{\rm ISCO}/r_g$ vs. $a_*$ relation).  Luckily, the techniques needed for measuring distance \citep{reid11}, mass, and inclination (\citealt{orosz11b,cantrell10}) for binary systems are well developed, and to date have been successfully applied to more than half a dozen black hole binary systems \citep{mcclintock11}.  

The continuum fitting method has evolved through a sequence of progressively more complex disc models.  The simplest model assumes multitemperature blackbody emission from the disc \citep{mitsuda84}.  Including relativistic effects such as Doppler beaming, gravitational redshifting, and light bending yielded the next generation of models (called \textsc{kerrbb} -- \citealt{li05}).  The current generation of disc models (named \textsc{bhspec} -- \citealt{davis05}; \citealt{davishubeny06}) frees itself from the blackbody assumption, and instead obtains the disc spectra by means of radiative transfer calculations.  All of the above mentioned models use the prescription of \citet[][hereafter NT]{NT} as the underlying disc model\footnote{This is the relativistic generalization of the standard \citet{shakura73} model for viscous geometrically thin but optically thick accretion discs.}, which assumes a perfectly dark void inside the ISCO.  The advance that we make in this work is to apply radiative transfer calculations to compute the spectrum for a general relativistic magneto-hydrodynamic (GRMHD) simulated disc that produces emission from inside the ISCO.  

\begin{figure}
\begin{center}  
\includegraphics[width=0.5\textwidth]{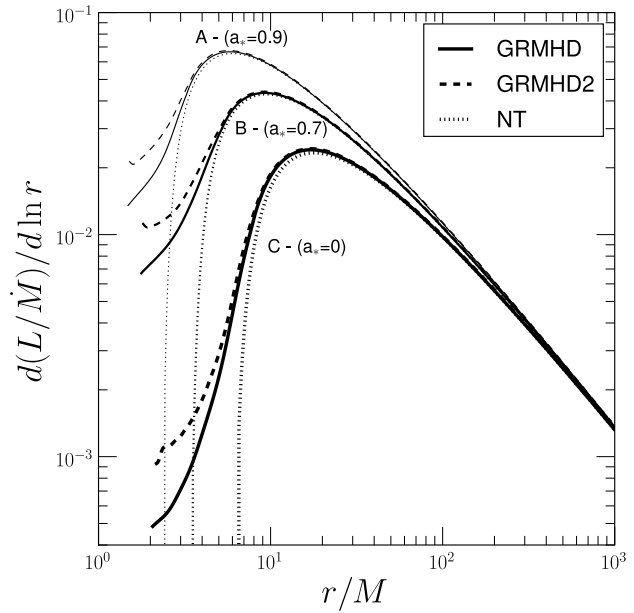}
\caption{Comparison of dimensionless luminosity profiles from the GRMHD simulations of Penna et. al 2011 (solid) with the standard disc model (dotted -- \citealt{NT,pagethorne74}).  Results from models A,B,C (defined in Table \ref{tab:discpar}), are shown.  The dashed line depicts an alternate measure of the GRMHD luminosity (described in \S\ref{sec:advection} and Appendix \ref{app:cooling}). Note that the luminosity in the standard \citetalias{NT} disc model goes to zero at the ISCO.\label{fig:lum}}  
\end{center}  
\end{figure}  

Currently, a crucial assumption in the continuum fitting enterprise is the perfect darkness of the plunging zone.  This scenario applies only to the idealized case of a razor thin unmagnetized disc (see \citetalias{NT}).  For finite thickness discs, it has been argued that as long as the disc remains geometrically thin (with disc aspect ratios $h/r \ll 1$), the disc can be well approximated by the NT model (\citealt{paczynski00,afshordi03,shafee08a}).  However, recent work with magnetized discs (\citealt{shafee08b,noble09,noble10,noble11,penna10}) suggests that there may be  departures from NT even in the limit of thin discs.  An observational difference between these magnetized discs and the classic NT discs is the appearance of non-negligible luminosity inside the plunging region (see Fig. \ref{fig:lum} for the result from \citet{penna10}, where the extra plunging region luminosity constitutes $\sim4$ per cent of the total).  Recent models of unmagnetized discs that include the physics of energy advection (\citealt{abram88}; most recently \citealt{sadowski09}, \citealt{sadowski11}) also share the feature of nonzero luminosity inside the ISCO, but in these advective models, the extra light only becomes significant when the accretion rate approaches the Eddington rate.

The primary goal of the present project is to \emph{investigate the importance of the neglected light from the inner plunging region of accretion discs}.  Since GRMHD discs exhibit the phenomenon that we wish to investigate (i.e. nonzero light from the plunging region), we use them as the basis for all our investigations.  We analyse the GRMHD simulations of \citet{penna10} in this work.  To connect the GRMHD simulations with observables, we compute realistic (i.e. radiative transfer based) disc spectra from them.  Essentially, we adapt the ideas pioneered in \textsc{bhspec} \citep{davis05,davishubeny06} to the domain of GRMHD discs.  One limitation of the GRMHD simulations is that they do not inherently model the radiation field of the disc fluid. The way we put the photons back in (for the purposes of computing disc spectra) is to do a radiative transfer post-processing step.  To get the integrated disc spectrum corresponding to a particular simulation run, we take the following three steps: 1) We slice the GRMHD disc into many individual annuli; 2) For each annulus, we compute the local spectrum through a radiative transfer calculation; 3) By means of ray tracing, we sum up the light from every annulus to get the overall spectrum of the accretion disc.

Several previous studies have also used GRMHD simulations to consider the impact of emission from the plunging region on disc spectra (e.g. \citealt{beckwith08,noble09,noble11,kulkarni11}), but all assumed (modified) blackbody emission.  Since the plunging region is precisely where the fluid becomes tenuous and optically thin, here the blackbody assumption is likely to be strongly violated.  Our work extends the analysis of \citet[][hereafter K2011]{kulkarni11}, which also made use of the GRMHD discs from \citet{penna10}.  The advance that we make beyond \citetalias{kulkarni11} is that we obtain disc spectra by means of radiative transfer calculations (this allows our work to be free from the blackbody assumption which underpins all previous work on GRMHD disc spectral modelling).

Our paper is organized as follows: In \S\ref{sec:GRMHDsim} we describe the setup and properties of the simulated GRMHD discs that we use to generate disc spectra.  We then detail the physics of the radiative transfer calculation in \S\ref{sec:tlusty}, and list the simulation quantities needed to generate the spectrum for each annulus.  In \S\ref{sec:slicing}, we explain how these simulation quantities are extracted, and we give a brief overview of the ray tracing process in \S\ref{sec:raytracing}.  The implications from the additional plunging region light are discussed in \S\ref{sec:spinResults}, and in \S\ref{sec:discussion} we briefly touch upon the limitations of our disc model. We summarize the key results in \S\ref{sec:summary}.  Appendices \ref{app:lum}--\ref{app:cooling} expand upon the technical details of various aspects of this work.


\section{GRMHD Simulations}\label{sec:GRMHDsim}

The simulation numerically evolves the 3D GRMHD equations in the Kerr spacetime via the code HARM (\citealt{gammie03,mckinney06,mckinney09}).  The code works in dimensionless units ($G=c=k_B=1$), and we assume that the fluid follows an ideal gas equation of state $P_{\rm gas}=(\Gamma-1)U$ where $P_{\rm gas}$ is the gas pressure, and $U$ is the gas internal energy density.  We choose the adiabatic index to be $\Gamma=4/3$, which corresponds to an ultrarelativistic equation of state.  We start off the simulation with a torus of gas that is initially in pure hydrodynamic equilibrium (\citealt{deVilliers03,gammie03}) threaded by a weak ($100 < P_{\rm gas}/P_{\rm mag} < 1000$) 4-loop poloidal magnetic field structure (see \citealt{penna10} for details on the field topology). The torus is also set up such that its orbital spin axis is aligned with the spin axis of the black hole.\\

There are three primary degrees of freedom for the torus: 1) the initial gas entropy, 2) the initial magnetic field strength and geometry, and 3) the initial angular momentum profile.  The choice of initial gas entropy controls the overall thickness of the disc, wheras the choice of initial magnetic field strength/geometry controls the strength of the turbulence (\citealt{beckwith08b}). The angular momentum profile sets the radial extent of the starting torus.  We evolve the system in time via a conservative Godunov scheme, with an additional caveat that we cool the gas via the following prescription:
\begin{equation} \label{eq:cooling}
\frac{dU}{d\tau}=-U\frac{\ln{(K/K_i)}}{\tau_{\rm cool}},
\end{equation}
where $U$ is the internal energy of the gas, $K=P/\rho^\Gamma$ is the gas entropy constant, $K_i = P_i/\rho_i^\Gamma$ is the initial entropy constant, and we set $\tau_{\rm cool} = 2\pi/\Omega_k$ for the cooling time-scale ($\Omega_k = \sqrt{GM/r^3}$).  After a gas element heats up from dissipation, energy is removed according to Eq. \ref{eq:cooling} such that the gas returns to its initial entropy\footnote{The cooling function in Eq. \ref{eq:cooling} is not invoked when $K<K_i$ (i.e. the cooling prescription does not add any heat to the fluid)}(this acts to preserve the initial aspect ratio of the disc). In the absence of a full radiative transfer calculation, the cooling function is a substitute for the radiative energy loss that we expect from a geometrically thin, optically thick disc.\\

The initial conditions of the specific simulations that we consider are listed in Table \ref{tab:discpar}.  All models are run using a fixed spherical polar mesh with $N_r=256$ (radial cells), $N_\theta=64$ (poloidal cells), and $N_{\phi}=32$ (toroidal cells), except for model E, which has twice the number of poloidal and toroidal cells.  The cell spacing is chosen such that resolution is concentrated near the horizon and the disc midplane (see \citealt{penna10} for details).  We find that the increase in resolution from model D to E has the effect of increasing $\alpha$.  Models A-C correspond to thin discs and most of our reported results focus on these models.  Model F differs from the other runs in its initial magnetic field topology in that it uses a single poloidal loop configuration rather than the 4-loop model adopted in the other runs.

\begin{table}
\caption{GRMHD disc parameters}
\begin{center}

\begin{threeparttable}

\begin{tabular}{cccccc}
\hline
 Model & $a_*$ & Target h/r & $L/L_{\rm Edd}$ & $\alpha$ & Comment\\
\hline
    A & 0.9 & 0.05 & 0.35 & 0.22 & --\\
    B & 0.7 & 0.05 & 0.32 & 0.10 & --\\
    C & 0.0 & 0.05 & 0.37 & 0.04 & --\\
    D & 0.0 & 0.1 & 0.70 & 0.04 & --\\
    E & 0.0 & 0.1 & 0.71 & 0.08 & High Res.\\
    F & 0.0 & 0.05 & 0.36 & 0.03 & 1-Loop\\
\hline
\end{tabular}

\begin{tablenotes}
\item Note -- The BH mass was set to be $M=10M_\odot$.  $L/L_{\rm Edd}$ and $\alpha$ are derived quantities that are measured from the simulations (see \S\ref{sec:slicing}). $\;$\citetalias{kulkarni11} also used Models A, B, C, and F in their work.
\end{tablenotes}

\label{tab:discpar}
\end{threeparttable}

\end{center}
\end{table}


\section{Annuli Spectra}\label{sec:tlusty}

We obtain the overall GRMHD accretion disc spectrum by summing up the light contributions from all annuli that comprise the disc.  Most previous work on disc spectra assume a modified blackbody prescription for the local spectrum emitted by each annulus. The blackbody is modified such that the specific intensity is given by:
\begin{equation}\label{eq:modBB}
I_{\nu}(T) = f^{-4} B_{\nu}(fT),
\end{equation} 
where $B_{\nu}(T)$ is the standard Planck function and $f$ is spectral hardening constant (often assumed to be 1.7 for X-ray binaries).  Following \citet{davishubeny06}, we improve on this assumption by solving the radiative transfer equation.  The emergent spectrum for each annulus is obtained through the stellar atmospheres code \textsc{tlusty} \citep{hubeny95}.  The code simultaneously solves for the vertical structure of a plane-parallel atmosphere and its angle dependent radiation field, and is \emph{independent of any blackbody assumption}.  The atmosphere is modelled as a series of 1D vertical cells that are in radiative, hydrostatic, and statistical equilibrium (which allows for departures from local thermodynamic equilibrium in the atomic populations).  In each cell, we obtain: the spectrum as a function of viewing angle $I_\nu(\theta)$, density $\rho$, gas temperature $T$, vertical height above the midplane $z$, and the particle composition ($n_e, n_{_{\rm HI}},n_{_{\rm HII}},{\rm etc}...$).

We are primarily interested in each annulus' emergent spectrum (i.e. $I_\nu(\theta)$ for the surface cell).  To obtain a unique solution in the stellar atmospheres problem, we need to specify the following 3 boundary conditions: 

\begin{enumerate}
\renewcommand{\theenumi}{(\arabic{enumi})}
\item The radiative cooling flux $F=\sigma_{\rm SB} T_{{\rm eff}}^4$ (where $T_{\rm eff}$ is the annulus effective temperature).\\
\item The vertical tidal gravity experienced by the fluid (parametrized by $Q=g_\perp/z$ where $g_\perp$ is the vertical acceleration and $z$ is the height above the midplane).\\
\item The column density to the midplane $m=\Sigma/2$ (where $\Sigma$ is the total column density of the annulus).
\end{enumerate}

The problem of computing the disc spectrum simply becomes one of obtaining the radial profiles of $T_{\rm eff}(r),\,Q(r)$, and $m(r)$ for the GRMHD disc.  We then compute the spectrum for each disc annulus with \textsc{tlusty} using the corresponding values of $T_{\rm eff}, \,Q$, and $m$.

\subsection{Assumptions in the TLUSTY model}
In our annulus problem, we make the assumption of uniform viscous heating per unit mass, and we ignore the effects of magnetic pressure support, fluid convection, and heat conduction.  We allow for deviations from LTE by explicitly computing the ion populations for H, He, C, N, O, Ne, Mg, Si, S, Ar, Ca, and Fe assuming solar abundances.  In the non-LTE calculations, only the lowest energy level is considered for each ionization state, except for the cases of H and He$^+$, which are modelled with 9 and 4 levels respectively.  The treatment of bound-free transitions includes all outer shell ionization processes, collisional excitations, and an approximate treatment for Auger (inner shell) processes.  Free-bound processes are modelled including radiative, dielectric, and three-body recombination. Free-free transitions are modelled for all listed elements, whereas bound-bound transitions are not modelled.  Comptonization is included through an angle-averaged Kompaneets treatment \citep{hubeny01}.  Finally, to reduce the number of independent vertical cells in the problem, we assume that the annulus is symmetric about the midplane.

One drawback of the algorithm used in \textsc{tlusty} (lambda-iteration and complete linearization, see \citealt{hubeny95} for details) is that \textsc{tlusty} requires an initial guess for the conditions in each vertical cell.  The code often fails to converge when the initial guess is poor.  However, if \textsc{tlusty} finds a converged solution, it is robust to the choice of initial guess.  Often, manual intervention (in the form of providing better initial guesses) is needed to ensure convergence, which means that the process of generating annuli spectra cannot be completely automated.  However, since only three parameters are necessary to uniquely specify an annulus, we have simply manually precomputed a grid of annuli spanning the full range of parameter space for the case of stellar mass black hole systems.  The grid spans $\log_{\rm 10}{T_{\rm eff}}\in\{5.5,\,5.6,\,\dots,\,7.3\}$, $\log_{\rm 10}{Q}\in\{-4.0,\,-3.9,\,\dots,\,9.0\}$, and $\log_{\rm 10}{m}\in\{0.5,\,0.6,\dots,\,2.8,\,2.9,\,3.0,\,4.0,\,5.0, \,6.0\}$.  We then interpolate on this grid to obtain any needed annuli spectra (see Appendix \ref{app:interpolation} for details on the interpolation process).


\section{Slicing the GRMHD disc into Annuli}\label{sec:slicing}

Since we wish to describe the accretion disc from the 3D GRMHD simulations as a series of annuli, we take out the poloidal and toroidal structure through an averaging process.  All quantities are first subject to azimuthal averaging (since we model the annuli as azimuthally symmetric structures), followed by vertical averaging (i.e. averaging over $\theta$) weighted by rest mass density.  For each annulus in the GRMHD simulation, we must identify $T_{\rm eff}(r), Q(r)$, and $\Sigma(r)$ before we can call on \textsc{tlusty} to provide the local spectrum. 

We obtain $T_{\rm eff}(r)$ directly from the simulation luminosity profile (\S\ref{sec:Teff}), $Q(r)$ from the Kerr metric and the fluid velocities (\S\ref{sec:Q}), and $\Sigma(r)$ from the fluid velocities and accretion rate (\S\ref{sec:sigma}).  Since the simulations are dimensionless, we must first choose a black hole mass $M$ and a disc luminosity $L/L_{\rm Edd}$ to dimensionalize the radial profiles of $T_{\rm eff}(r)$, $Q(r)$, and $\Sigma(r)$.  For all the simulation runs, we have chosen a fiducial value of $M=10\,{\rm M}_{\odot}$, and we set $L/L_{\rm Edd}$ so the disc thickness in the radiative model matches that of the GRMHD simulation (see \S\ref{sec:thicknessMatching}).  

One complicating factor is that the simulated disc is only reliable for a short range of radii.  Beyond a certain critical radius  $r > r_{ie}$ (where $r_{ie}$ stands for the radius of inflow equilibrium -- see \citealt{penna10} for a detailed discussion), the simulation has not been run long enough for the fluid to reach its equilibrium configuration.  The relevant physical time-scale for reaching inflow equilibrium is the accretion time $t_{\rm acc}=r/u^r$, where $u^r$ is the fluid's radial velocity.  The problem with fluid outside of $r_{ie}$ is that this fluid still has memory of its initial conditions (i.e. fluid outside of $r_{ie}$ has $t_{\rm acc} > t_{\rm sim}$, where $t_{\rm sim}$ is the total runtime of the simulation).  Beyond the inflow equilibrium point, we instead turn to an analytic disc model that is matched to the simulation disc (we use a generalized NT model that allows for nonzero stresses at the ISCO, see Appendices \ref{app:lum} and \ref{app:vr} for details).  The analytic model also requires us to specify a disc viscosity parameter $\alpha$ (defined as the ratio between the vertically integrated stress and pressure, see \citealt{NT}).  Table \ref{tab:discpar} lists the values of $L/L_{\rm Edd}$ and $\alpha$ corresponding to the different simulation runs, where both parameters are determined by matching the GRMHD disc at $r_{ie}$ to the analytic generalized NT disc (see \S\ref{sec:thicknessMatching} and \S\ref{sec:vrMatching}).

\subsection{Flux profile - \texorpdfstring{$T_{\rm eff}(r)$}{Teff}}\label{sec:Teff}

From the simulations, we have extracted in the Boyer-Lindquist ($t,r,\theta,\phi$) frame, the fluid four-velocity  ($u^t, u^r, u^\phi, u^\theta$), and luminosity $L_{BL}$, where the luminosity is computed from the cooling function $dU/d\tau$ (see Eq. \ref{eq:cooling}) in the following fashion:
\begin{equation} \label{eq:lum}
L_{\rm BL}(r) = \frac{1}{t_f-t_i} \int\limits_{\phi=0}^{2\pi} \int\limits_{\theta=0}^{\pi} \int\limits_{r'=r_H}^{r} \int\limits_{t=t_i}^{t_f} \frac{dU}{d\tau}u_{t}\sqrt{-g} dt dr' d\theta d\phi,
\end{equation}
where $t_i$ and $t_f$ are the initial and final times considered for the time integral, $r_H$ is the black hole horizon, $u_t$ is the fluid's conserved specific energy, and $\sqrt{-g}$ is the determinant of the Kerr metric.  We do the integration for only the bound fluid, and then obtain the comoving flux by transforming from the Boyer-Lindquist flux (see \citetalias{kulkarni11}):
\begin{equation} \label{eq:flux}
\sigma_{\rm SB} T_{\rm eff}^4(r) = F_{\rm com}(r) = \frac{F_{\rm BL}(r)}{-u_t} = \frac{1}{-4\pi r u_t} \frac{dL_{\rm BL}(r)}{dr}.
\end{equation}
However, since the simulations are converged only for a short radial extent, we extrapolate the flux profile to large radii by matching the GRMHD flux with an analytic disc model at a matching radius $r_{ie}$.   The extrapolation model we use is the same as \citetalias{kulkarni11}, which is a generalized \citet{pagethorne74} model (see Appendix \ref{app:lum}) that does not assume zero torque at the disc inner boundary, and hence allows for nonzero ISCO luminosities. 

\begin{figure}
\begin{center}  
\includegraphics[width=0.5\textwidth]{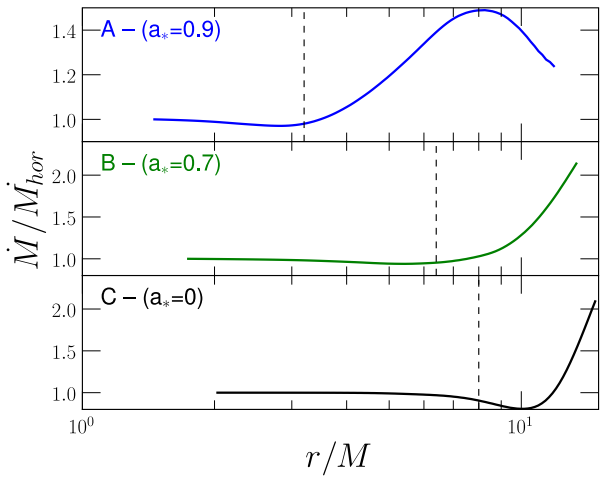}  
\caption{The GRMHD simulation accretion rates normalized to $\dot{M}$ at the horizon are shown.  We identify the region left of the vertical dashed line as converged, so the line marks the boundary $r_{ie}$ where we match the GRMHD disc to an analytic model. \label{fig:Mdot}}  
\end{center}  
\end{figure}

To determine the point at which we can no longer trust the simulations (in other words, to locate $r_{ie}$), we look at the radial profile of the GRMHD accretion rate (Figure \ref{fig:Mdot}).  Beyond a certain radius, we see that the accretion rate begins to significantly deviate from a constant value (the deviation from a constant accretion rate only occurs in fluid that has not reached inflow equilibrium).  We pick the matching point to be the outermost point where we have well behaved accretion, though the final luminosity profile is fairly insensitive\footnote{The insensitivity of the luminosity extrapolation to the matching radius $r_{ie}$ was demonstrated in appendix A of K2011.} to our choice of matching radius (i.e. after varying the matching location by $\pm20$ per cent, the luminosity profile changes by less than 10 per cent).

\subsubsection{Disc thickness matching (\texorpdfstring{$L/L_{\rm Edd}$)}{LEdd}}\label{sec:thicknessMatching}

\begin{figure}
\begin{center}  
\includegraphics[width=0.5\textwidth]{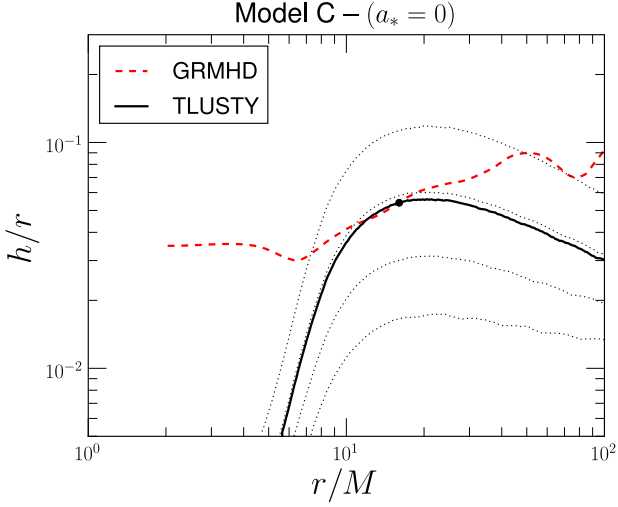}
\caption{Shown are the $h/r$ disc thickness profiles for model C, as computed by \textsc{tlusty} and our GRMHD simulation, where $h=\int \rho |z| dz/\int \rho dz$.  The black dotted lines depict the \textsc{tlusty} disc thicknesses for different choices of luminosity (the four sets of dotted lines correspond to $L/L_{\rm Edd}=0.1,0.2,0.4,0.8$).  The black solid line denotes the matched thickness profile with $L/L_{\rm Edd}=0.37$.  The black dot represents the radius where the luminosity is greatest, which we have chosen to be the radius at which we match the \textsc{tlusty} and GRMHD thickness profiles. \label{fig:hrMatch}}  
\end{center}  
\end{figure}

The GRMHD luminosity that we measure (as shown in Figure \ref{fig:lum}) is actually in dimensionless units of $L/\dot{M}_{\rm sim}$ where $\dot{M}_{\rm sim}$ is the local accretion rate of the simulations (see Figure \ref{fig:Mdot}). To calculate $T_{\rm eff}$ in dimensioned units of [K], we need to first determine the dimensioned luminosity corresponding to the GRMHD simulations.  Unfortunately, the simulations do not include radiation physics, so the simulation fluid does not have a set density scale, which means there is no direct way of measuring $\dot{M}$ and hence $L$.  We thus resort to an indirect method for determining this accretion rate.  We identify $L/L_{\rm Edd}$ of the simulation with the value that matches the disc thickness in both the GRMHD simulated disc and the \textsc{tlusty} annuli\footnote{\citetalias{kulkarni11} also applied this method to dimensionalize their luminosity profiles} (see Figure \ref{fig:hrMatch}).  However, since \textsc{tlusty} ignores magnetic support and the simulations ignore radiation pressure, the \textsc{tlusty} based and GRMHD thickness profiles have completely different shapes.  We opt to match the thickness at the radius corresponding to the luminosity peak (see Figure \ref{fig:hrMatch}).

\begin{figure}  
\begin{center}  
\includegraphics[width=0.5\textwidth]{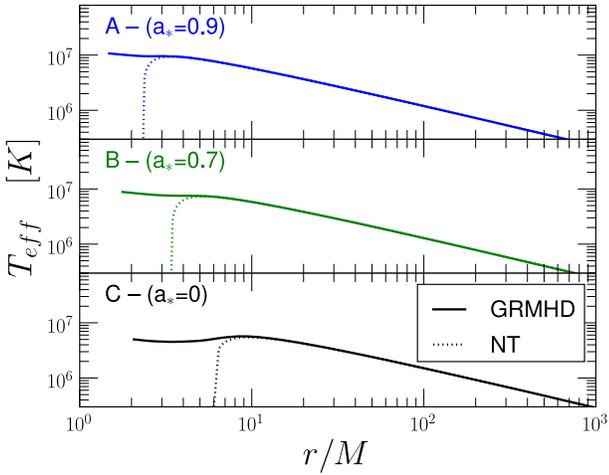} 
\caption{Here we plot the $T_{\rm eff}$ flux profile, computed from the luminosity profiles of Figure \ref{fig:lum}, for the accretion rates listed in Table \ref{tab:discpar}.\label{fig:Teff}}  
\end{center}  
\end{figure}

Although the luminosity profile drops off fairly rapidly inside the ISCO (see Fig. \ref{fig:lum}), we find that the disc's effective temperature remains roughly constant (Fig. \ref{fig:Teff}).  Therefore the falloff in disc luminosity when approaching the BH horizon is purely a geometric effect.  Annuli near the horizon have less surface area, and hence they contribute less to the total luminosity.

\subsection{Vertical gravity profile - \texorpdfstring{$Q(r)$}{Q(r)}}\label{sec:Q}

The vertical gravity parameter $g_\perp$, which represents the tidal vertical acceleration in the disc, is obtained by evaluating the $R^{\hat{z}}_{\hat{t}\hat{z}\hat{t}}$ component of the Riemann curvature tensor in the comoving frame.  A commonly used prescription for the vertical gravity is that of \citet{riffert95}, which assumes that the disc fluid moves along circular geodesics; this approach becomes invalid inside the plunging zone.  We thus turn to the prescription of \citet{abram97}, which relaxes the circular orbit assumption:
\begin{equation}
g_\perp(r,z) = \Omega_k^2 R_{z}(r) \cdot z,
\end{equation}
where $\Omega_k = (GM/r^3)^{1/2}$ is the Keplerian orbital frequency, and the dimensionless relativistic factor $R_{z}(r)$ is given by:
\begin{equation}\label{eq:abram}
R_z(r) = \left(\frac{u_\phi^2+a_*^2(u_t-1)}{r}\right).
\end{equation}
In Eq. \ref{eq:abram} we use the four-velocity corresponding to the midplane fluid from the GRMHD simulations for $r$$<$$r_{ie}$ (as determined from Figure \ref{fig:Mdot}).  For $r$$\,>\,$$r_{ie}$, we use circular geodesics as the four-velocity in Eq. \ref{eq:abram}.  We find that the two prescriptions for vertical gravity \citep{riffert95,abram97} give nearly identical results outside the ISCO.

For convenience, we define the function $Q(r)$ such that:
\begin{equation}\label{eq:Qdef}
g_\perp(r,z) = Q(r)\cdot z,
\end{equation}
and we interpret $Q(r)$ as the radial dependence of the vertical gravity.

\subsection{Column density profile - \texorpdfstring{$\Sigma(r)$}{Sig(r)}}\label{sec:sigma}

Since the GRMHD simulation does not include radiation physics, the simulation mass density is scale-free, so we cannot directly read out the disc column mass density in physical units.  We dimensionalize the simulation density by solving for $\Sigma$ in the vertically integrated mass conservation equation after having picked a constant accretion rate $\dot{M}$:\footnote{We set $\dot{M}$ to be the value corresponding to the disc luminosity as determined in \S\ref{sec:thicknessMatching}. They are related by $L=\eta \dot{M}c^2$ where $\eta(a_*)$ is the spin dependent accretion efficiency of the NT disc.}
\begin{equation}\label{eq:masscons}
\dot{M}=2\pi r \Sigma \tilde{u}^r,
\end{equation}
where $r$ is the BL radial coordinate, and $\tilde{u}^r$ is the mass-averaged radial velocity for the GRMHD simulations, computed via:
\begin{equation}\label{eq:ur}
\tilde{u}^r=\frac{\int_{\theta=0}^{\pi} \hat{\rho} u^r \sqrt{-g} d\theta}{\int_{\theta=0}^{\pi} \hat{\rho}\sqrt{-g} d\theta}. 
\end{equation}
\subsubsection{Radial velocity matching (\texorpdfstring{$\alpha$}{alpha})}\label{sec:vrMatching}
\begin{figure} 
\begin{center}  
\includegraphics[width=0.5\textwidth]{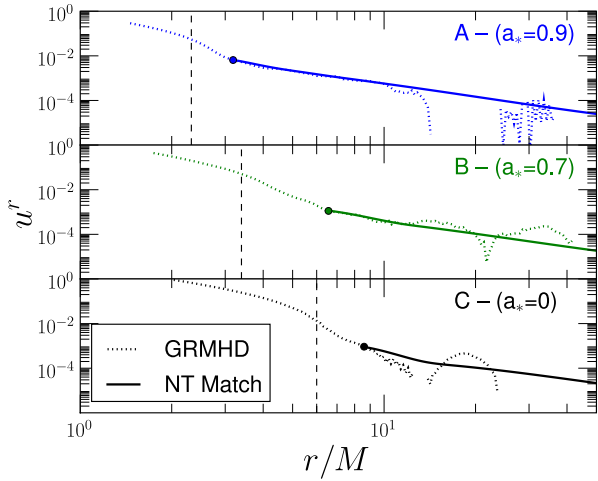} 
\caption{A comparison between the mass averaged simulation radial velocity profile (dotted) and the final matched radial velocity profile (solid line).  The matching point is depicted as the large coloured circle, whereas the vertical dashed line denotes the location of the ISCO.  The gaps in the GRMHD radial velocity profile represent grid cells whose radial velocity is pointed outwards.\label{fig:ur}}  
\end{center}  
\end{figure}

  At large radii ($r > r_{ie}$), the simulation is no longer converged, and we resort to matching the simulation result with an analytical disc model.  Rather than directly using the \citet{NT} disc model, we re-solve the NT disc equations using our matched GRMHD luminosity profile (see Appendix \ref{app:vr} for details).  The final matched radial velocity profile is shown in Figure \ref{fig:ur}.

\begin{figure}
\begin{center}  
\includegraphics[width=0.5\textwidth]{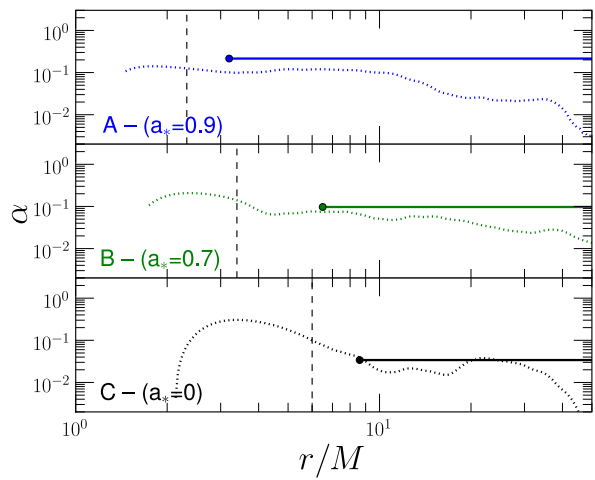} 
\caption{The vertically averaged $\alpha$ profile as computed from the GRMHD simulation (dotted) compared to the $\alpha$ obtained by the radial velocity matching method (solid horizontal lines, corresponding to the $\alpha$ values listed in Table \ref{tab:discpar}).  For reference, we have plotted the ISCO locations using dashed vertical lines.\label{fig:alpha}}  
\end{center}  
\end{figure}

We choose the disc viscosity $\alpha$ so as to ensure continuity in the radial velocity profile (Fig. \ref{fig:ur}).  The matching values of $\alpha$ for each disc are listed in Table \ref{tab:discpar}.   The location of the radial velocity matching point is set to be the same as the luminosity matching point.  To verify that our $\alpha$ value from radial velocity matching is sensible, we compare it to the directly computed value from the simulation $\alpha_{\rm sim}=\tau_{\hat{r}\hat{\phi}}/(hP_{\rm tot})$ where $\tau_{\hat{r}\hat{\phi}}$ is the height integrated shear stress, and $hP_{\rm tot}$ is the height integrated total pressure (where $P_{\rm tot}=P_{\rm mag}+P_{\rm gas}$).  The near continuity in the simulation $\alpha$ profile with the matched value suggests that our method for determining $\alpha$ is sound.

\begin{figure}
\begin{center}  
\includegraphics[width=0.5\textwidth]{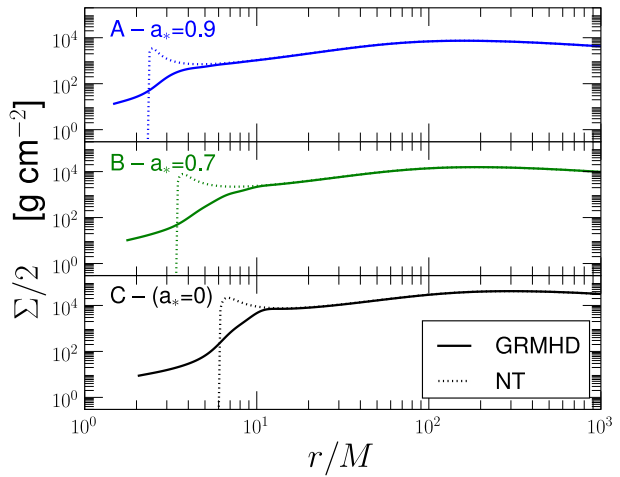} 
\caption{Dimensioned column density profiles for the GRMHD discs (solid lines) compared to \citetalias{NT} discs (dotted lines).  Note the sharp drop in NT column density as the fluid reaches the plunging region, in contrast to the more gradual tapering observed in the GRMHD simulations.\label{fig:m}}  
\end{center}  
\end{figure}

From the radial velocity profile and from our choice of constant $\dot{M}$, the mass continuity equation (Eq. \ref{eq:masscons}) yields the final matched disc column density profile (Fig. \ref{fig:m}).

\section{Ray Tracing}\label{sec:raytracing}

The final step in the process of generating the accretion disc's SED is to sum the light from all the individual annuli.  The observed disc image is determined by shooting a series of parallel light rays from an image plane that is perpendicular to the black hole line of sight.  We partition the image plane into a sequence of polar grid cells with logarithmic radial spacing (to concentrate resolution near the black hole), and with uniform angular spacing.  To produce the ray traced images, we shoot rays for $N_r=300$ radial grid cells and $N_\phi=100$ polar cells.  The grid is then further squashed by a factor of $\cos i$ to match the geometry of the inclined disc.

\begin{figure*}
\begin{center}  
\includegraphics[width=1.0\textwidth]{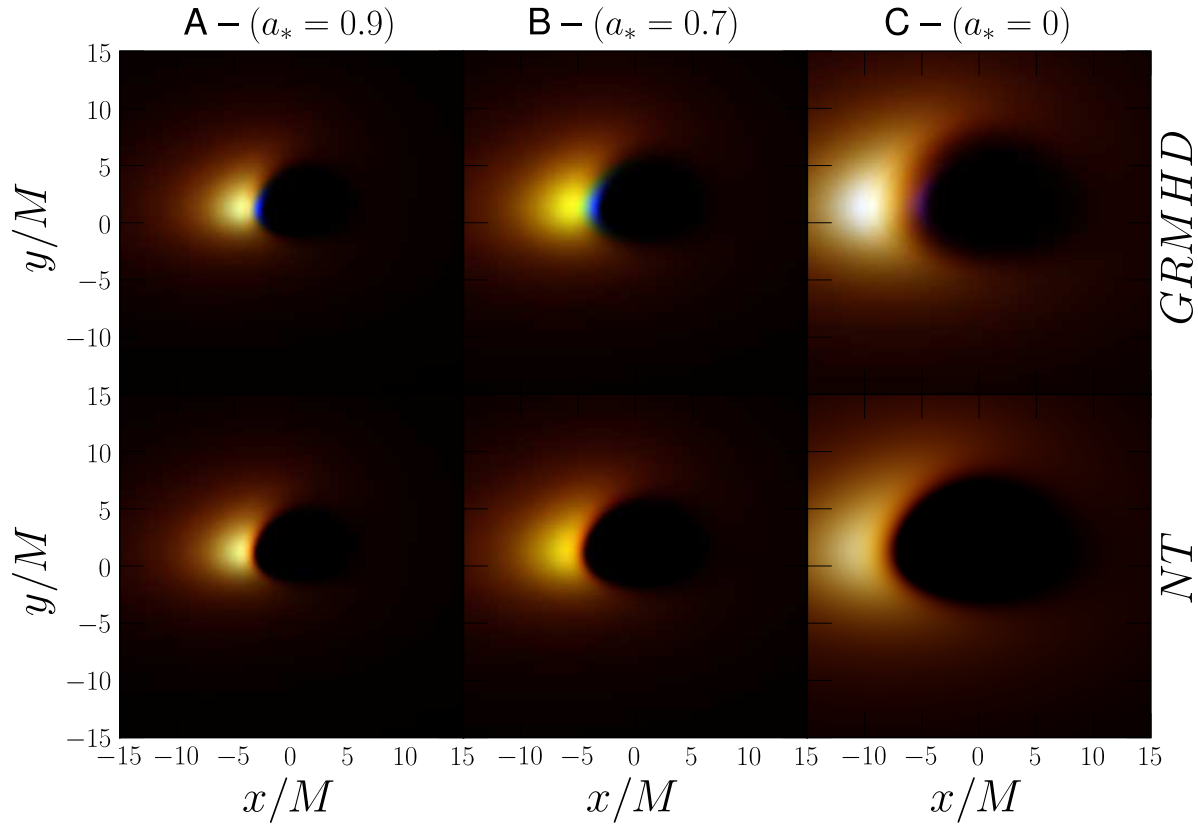}  
\caption{Colour images of the inner disc ($r<15M$) produced via ray tracing (viewed at an inclination angle of $i=60^\circ$) for models A, B, and C in Table \ref{tab:discpar}.  The colours correspond to the flux $F_\nu$ integrated over different energy bands, where red colour corresponds to the energy band $E<4\,{\rm keV}$, green $4\,{\rm keV}<E<12\,{\rm keV}$, and blue $E>12\,{\rm keV}$.  For each value of BH spin, the colour mapping for the GRMHD and NT panels are identical, and each colour channel is normalized to the maximum flux in that channel.  Note the appearance of a blue spot in the GRMHD disc plunging region.  This blue spot is responsible for the appearance of a nonthermal high energy power-law feature in the disc spectra (see Figure \ref{fig:spectraPlot}).\label{fig:raytraceImages}}  
\end{center}  
\end{figure*}

A single ray is shot for each grid cell, and by numerically integrating the (second-order) geodesic equations for light:
\begin{equation}
\frac{d^2x^\alpha}{d\lambda^2}+\Gamma^{\alpha}_{\beta\gamma}\frac{dx^\beta}{d\lambda}\frac{dx^\gamma}{\lambda}=0,
\end{equation}
 where $\lambda$ is an affine parameter along the geodesic, and $\Gamma^{\alpha}_{\beta\gamma}$ are the connection coefficients, we locate the first disc midplane intersection for each light ray (we stop at the disc midplane since we have an optically thick disc -- see Figure \ref{fig:tauEff} for a plot of the optical depth profile).  For simplicity, we opt to use the disc midplane as the point of light emission rather than using each annulus' precise photosphere.  \citetalias{kulkarni11} suggested that the latter is important only for thick discs viewed at high inclinations angles (which is when disc self-shadowing becomes important).  In this work, we do not expect the distinction between midplane and photosphere to be important since we are dealing with thin discs (where $h/r \ll \cos i$).  The ray tracing code used in this work was originally developed by \citet{scherbakov11}, with further refinements by \citetalias{kulkarni11}.

For each ray, we compute (by interpolation on the GRMHD grid cells, see Appendix \ref{app:interpolation} for details) the local values of: flux $T_{\rm eff}$, column density $\Sigma$, vertical gravity $Q$, and comoving light ray incidence angle $\mu$.  The local comoving spectrum is then obtained by interpolating on our grid of \textsc{tlusty} atmosphere models.  We then apply the relevant Doppler boosting and gravitational redshifting to these spectra (see Fig. \ref{fig:raytraceImages} for the ray traced images).  Finally, the overall disc spectrum is obtained by integrating the spectra corresponding to these light rays over the apparent disc area in the image plane.  We take a fiducial BH distance of 10\,kpc when generating the disc spectra.


\section{Results}\label{sec:spinResults}

Qualitatively, we can spot a few differences in Figure \ref{fig:raytraceImages} between the GRMHD (top) and NT (bottom) panels.  We see that for spinning black holes, the plunging fluid emits in the hard X-rays (appearing as a $>12\,{\rm keV}$ blue smudge in Fig. \ref{fig:raytraceImages}).  Even outside the plunging region, there is a noticeable increase in disc luminosity, which results in harder annuli spectra everywhere (compare the bright white Doppler spot in the upper right (Model C, GRMHD) panel, with the duller orange spot in the lower right (Model C, NT) panel.

\begin{figure}
\begin{center}  
\includegraphics[width=0.5\textwidth]{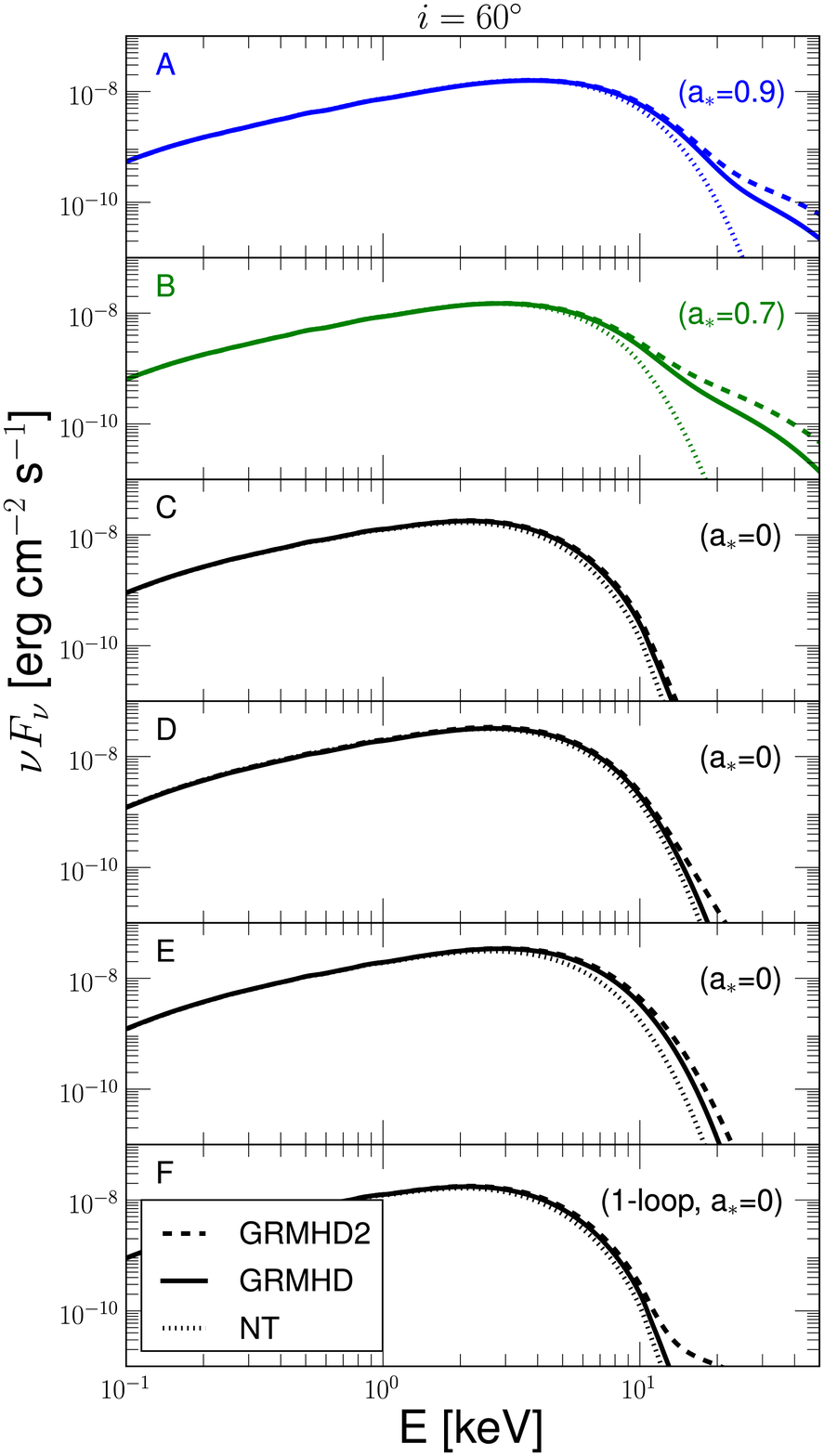} 
\caption{GRMHD and \citetalias{NT} disc spectra for each of the 6 models listed in Table \ref{tab:discpar}.  The flux normalization corresponds to an assumed distance of 10\,kpc.  Note that the GRMHD spectra peak at slightly higher energies than their NT counterparts.  Note also the emergence of a high energy power-law tail for the spinning black holes.  The dashed spectra (labelled GRMHD2) correspond to the alternate GRMHD luminosity profile (dashed line in Figure \ref{fig:lum}), discussed in \S\ref{sec:advection} and Appendix \ref{app:cooling}. }\label{fig:spectraPlot}
\end{center}  
\end{figure}

By integrating the flux in the image plane, we arrive at the observed disc spectrum (see Fig. \ref{fig:spectraPlot}).  For spinning black holes, the spectra corresponding to the GRMHD simulations appear to exhibit a power-law component at high energies.  We discuss this effect in \S\ref{sec:powerlaw}.  In addition, for all models, the location of the energy peak in the spectra for the GRMHD discs is shifted towards higher energies.  The harder overall spectrum of GRMHD compared to NT implies that there would be an asymmetric error in BH spin estimates (since the current fitting models are based on the cooler \citetalias{NT} disc).  The quantitative size of this spin bias is discussed in \S\ref{sec:spinbias}.

\subsection{Power law tail}\label{sec:powerlaw}

Observationally, high-energy power-law tails are often seen in the X-ray spectra of black hole binary systems (e.g. \citealt{miyamoto91,mcclintock01,reis10}).  The current leading theory to explain these high-energy power-law phenomena is the action of a hot optically thin corona \citep{zhang97b}, which produces the power-law tail by means of either Comptonization of disc photons, synchrotron radiation, and/or synchrotron self-Comptonization (\citealt{miller07,mcclintockremillard06}).  Despite the many proposed coronal models (e.g. \citealt{haardt91,dove97,kawaguchi00,liu02}), there is no agreed-upon standard (i.e. there is scarce agreement on the geometry and physical properties of the corona).  Although a hot corona is often invoked to explain the existence of a high energy power-law tail, in this work we naturally recover a hot power-law tail from just the thermal disc fluid that occupies the plunging region.

\begin{figure}
\begin{center}  
\includegraphics[width=0.5\textwidth]{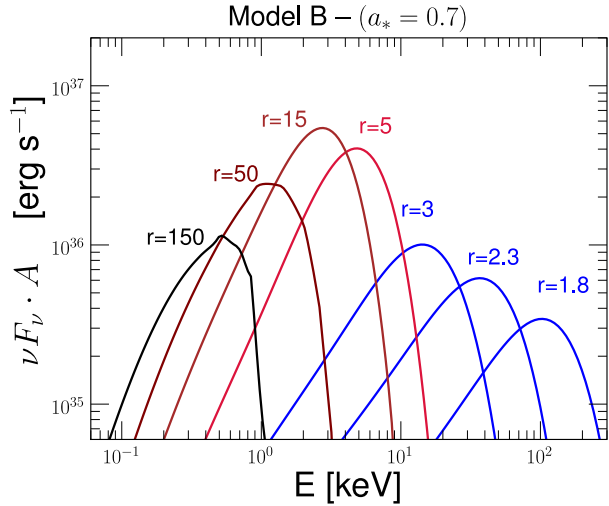} 
\caption{Local emergent fluxes from various annuli for model B ($a_*=0.7$), weighted by their face-on emitting areas (i.e. $A=2\pi r \Delta r$, where $\Delta r$ is the radial width of the annulus).  Plunging region annuli are coloured blue, whereas annuli beyond the ISCO are coloured red to black (for reference, the ISCO is located at $3.4M$, and the horizon is at $1.7M$).  Doppler and gravitational redshift corrections have not been applied to these spectra. \label{fig:annuliSpectra}}  
\end{center}  
\end{figure}

To better understand the origin of the apparent high-energy power-law tail in our models, we examine the area-weighted local spectrum for several annuli in model B ($a_*=0.7$), chosen since this simulated disc exhibits the strongest tail relative to the disc continuum.  From Figure \ref{fig:annuliSpectra}, we see that the high energy power-law tail stems solely from the emission of annuli inside the plunging region (coloured in blue).  Furthermore, we see that the power-law emerges from the combined emission of plunging annuli at various locations (i.e. the envelope of spectra from $r=2M - 3M$ comprise the power-law).  No single annulus emits the full power-law.

In our plunging region model, we find that the strength of the high energy power-law also depends on BH spin.  The spinless models (C, D, E, F) do not appear to produce any significant power-law component.  The reason for why model B ($a_*=0.7$) produces a stronger power-law than model A ($a_*=0.9$) is simply that the plunging region in B covers a larger area in the image plane than A, leading to a larger overall plunging region flux (compare the relative sizes of the blue plunging region blobs in Figure \ref{fig:raytraceImages}).  We also find a correlation in the strength of the power-law with disc viewing angle;  edge-on discs produce the strongest power laws relative to the thermal continuum since Doppler beaming increases with inclination angle, and hence acts to preferentially boost the plunging region emission.

\begin{figure}
\begin{center}  
\includegraphics[width=0.5\textwidth]{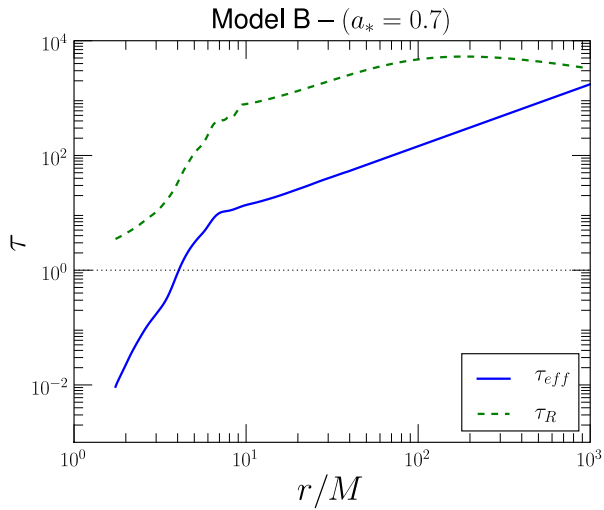} 
\caption{Optical depths corresponding to model B ($a_*=0.7$). The total optical depth $\tau_R = \Sigma \kappa_R$ is computed using the Rosseland mean opacity $\kappa_R$ (including both absorption and scattering).  The effective optical depth is computed as $\tau_{\rm eff} = \sqrt{3\tau_{\rm abs}\tau_{\rm scat}}$ where $\tau_{\rm abs}$ and $\tau_{\rm scat}$ are the individual absorption and scattering optical depths respectively.\label{fig:tauEff}}  
\end{center}  
\end{figure}

We find that the spectral hardening factor for the plunging annuli appears to start growing non-linearly as the gas approaches the horizon (see Fig. \ref{fig:annuliSpectra} and note the rapid shift in the spectral peak locations for the blue plunging region annuli, despite that $T_{\rm eff}$ is nearly constant in the plunging zone as per Fig. \ref{fig:Teff}).  The physical reason for this rapid spectral hardening is simply that the plunging gas quickly becomes extremely hot.  Upon entering the plunging region, the fast radial plunge causes the annuli to rapidly thin out in column density.  The dropping column mass allows the annuli to make the transition from being optically thick to effectively optically thin\footnote{An effectively optically thin medium is one that is optically thin to absorption, and optically thick to scattering.} (see Fig. \ref{fig:tauEff}).  Despite this drop in column density, the gas needs to maintain a roughly constant cooling rate (i.e. $T_{\rm eff}$ remains constant in the plunging zone -- see Fig. \ref{fig:Teff}).  A property of effectively optically thin gas is that it cannot radiate efficiently \citep{shapiro76}.  To keep a constant cooling rate despite the rapidly dropping optical depth, the gas must heat up tremendously, leading to very hot emission.

The power laws produced by the plunging region in our models have photon power-law indices that range from $\Gamma = 4-6$, with scattering fractions\footnote{The scattering fraction denotes the fraction of thermal seed photons that undergo Comptonization to produce the power-law component.} ranging from 1-15 per cent.  In our model, we find that the scattering fraction (and hence strength of the power-law) increases monotonically with inclination angle, whereas the photon power-law index decreases monotonically.  The scattering fraction range spanned by our models is comparable to the observed range covered by the thermal-dominant state, as well as much of the range of the intermediate, and steep power-law states for black hole binaries (which ranges from 0-25 per cent -- see \citealt{steiner09b,steiner11}).  However, even in the steep power-law state, the observed photon index range of $\Gamma = 2.2-2.7$\citep{gou09,gou11,steiner11} is much less steep than what we obtain from the plunging region of our models.  This discrepancy may simply be due to a limitation of our model, which neglects the physics of disc self-irradiation.  Including this effect would result in a hotter disc that would act to boost the strength of the high energy tails.  Also, self-irradiation becomes increasingly important as one approaches the black hole (where light bending is most severe).  This implies that including irradiation will likely result in power laws that are less steep (since the emission from the hottest, innermost annuli will be boosted the most by this self-irradiation effect).  We speculate that a more complete treatment of the plunging region (that includes the physics of disc irradiation and magnetic fields) would produce high energy tails that are better matched to observations.

Since many of our approximations become poor in the plunging region (see \S \ref{sec:discussion} for a discussion of some of the pitfalls of our method), we strongly caution against reading too much into the quantitative details of the power laws of Figure \ref{fig:spectraPlot} (i.e. slope, normalization).  The high energy power-law result should be taken only at the qualitative level; although the plunging region luminosity is dwarfed by the disc luminosity, it starts to dominate the flux at $E>20\,$keV taking on a nonthermal shape that appears as a power-law.

\subsection{Quantitative effect on spin}\label{sec:spinbias}

As already discussed, the thermal emission from the disc is slightly stronger and hotter in the GRMHD models compared to the equivalent NT model.  This will introduce an error in BH spin estimates obtained from the continuum fitting method.  To quantify the effect, we generate mock observations of our GRMHD discs and we fit these simulated observations with the currently used suite of continuum-fitting disc models.  For simplicity, to generate these mock observations, we use an idealized instrument with effective area that is independent of photon energy.  We choose the instrument's effective area\footnote{This is modelled after the Rossi X-Ray Timing Explorer's total effective area in the hard energy band ($>20\,$keV), see \citet{jahoda06}} to be $1000\,{\rm cm}^2$, and use a 3h exposure time.  We compute photon statistics in 1000 energy channels uniformly spaced in $\log E$, ranging from $0.4\,{\rm keV} < E < 50.0 \,{\rm keV}$.

We then fit our simulated disc observations with \textsc{bhspec} (\citealt{davis05,davishubeny06}), which is a set of (TLUSTY based) disc spectra often used in continuum fitting.  The \textsc{bhspec} spectra are computed in much the same way as our procedure in \S\ref{sec:intro}, except that \textsc{bhspec} uses a classic \citetalias{NT} disc instead of our GRMHD disc.  Whenever necessary, we also attempt to fit a power-law tail to the models using \textsc{simpl} \citep{steiner09a}, which is a physically motivated Comptonization model that has been quite successful in fitting many observed high energy power-law tails \citep{steiner09b}.

\begin{table*}
\caption{The recovered BH spins from XSPEC fitting of simulated disc observations.}

\begin{center}

\begin{threeparttable}
\begin{tabular}{cccccc}

\hline
$i(^\circ)$ & NT & GRMHD & GRMHD+PL & GRMHD2$^\dagger$ & GRMHD2+PL$^\dagger$ \\
\hline
\hline
\multicolumn{6}{c}{(Model A -- $a_*=0.9, \; L/L_{\rm Edd} = 0.35, \; \alpha=0.22$) }\\
\\
    15 & 0.902$\pm$0.004 & 0.909$\pm$0.018 &            --   & 0.913 $\pm$0.017 & --\\
    30 & 0.909$\pm$0.005 & 0.914$\pm$0.015 &            --   & 0.916 $\pm$0.016 & --\\
    45 & 0.904$\pm$0.006 & 0.912$\pm$0.013 &            --   & 0.916 $\pm$0.013 & 0.916$\pm$0.005\\
    60 & 0.902$\pm$0.005 & 0.919$\pm$0.012 & 0.913$\pm$0.006 & 0.921 $\pm$0.011 & 0.911$\pm$0.004\\
    75 & 0.897$\pm$0.007 & 0.925$\pm$0.011 & 0.916$\pm$0.005 & 0.934 $\pm$0.010 & 0.908$\pm$0.004\\
    \\
\hline 
\multicolumn{6}{c}{(Model B -- $a_*=0.7, \; L/L_{\rm Edd} = 0.32, \; \alpha=0.10$)} \\
\\
    15 & 0.68$\pm$0.01 & 0.73$\pm$0.03 & 0.72$\pm$0.03 & 0.75$\pm$0.03 & 0.73$\pm$0.03\\
    30 & 0.69$\pm$0.01 & 0.74$\pm$0.03 & 0.72$\pm$0.02 & 0.75$\pm$0.03 & 0.72$\pm$0.03\\
    45 & 0.69$\pm$0.01 & 0.75$\pm$0.02 & 0.71$\pm$0.03 & 0.76$\pm$0.03 & 0.72$\pm$0.03\\
    60 & 0.70$\pm$0.01 & 0.75$\pm$0.02 & 0.71$\pm$0.03 & 0.76$\pm$0.02 & 0.71$\pm$0.04\\
    75 & 0.70$\pm$0.01 & 0.76$\pm$0.01 & 0.69$\pm$0.04 & 0.77$\pm$0.03 & 0.69$\pm$0.04\\
    \\
\hline    
\multicolumn{6}{c}{(Model C -- $a_*=0, \; L/L_{\rm Edd} = 0.37, \; \alpha=0.04$)} \\
\\
    15 & -0.12$\pm$0.04 & -0.04$\pm$0.09 & --  & -0.01$\pm$0.09 & --\\
    30 & -0.10$\pm$0.04 & -0.03$\pm$0.09 & --  & 0.01$\pm$0.09 & --\\
    45 & -0.09$\pm$0.04 & 0.01$\pm$0.08 & --  & 0.03$\pm$0.08 & --\\
    60 & -0.08$\pm$0.04 & 0.03$\pm$0.08 & --  & 0.07$\pm$0.09 & --\\
    75 & -0.06$\pm$0.05 & 0.05$\pm$0.09 & --  & 0.10$\pm$0.10 & --\\
    \\
\hline    
\multicolumn{6}{c}{(Model D -- $a_*=0, \; L/L_{\rm Edd} = 0.70, \; \alpha=0.04$)} \\
\\
    15 & -0.19$\pm$0.02 & -0.01$\pm$0.09 & --              &  0.04$\pm$0.09 & --\\
    30 & -0.17$\pm$0.02 &  0.00$\pm$0.09 & --              &  0.05$\pm$0.09 & --\\
    45 & -0.15$\pm$0.03 &  0.01$\pm$0.08 & --              &  0.08$\pm$0.09 & --\\
    60 & -0.14$\pm$0.03 &  0.03$\pm$0.08 & --              &  0.11$\pm$0.09 & 0.08$\pm$0.10\\
    75 & -0.12$\pm$0.04 &  0.08$\pm$0.09 &  0.04$\pm$0.09  &  0.17$\pm$0.09 & 0.07$\pm$0.13\\
    \\
\hline
\multicolumn{6}{c}{(Model E -- $a_*=0, \; L/L_{\rm Edd} = 0.71, \; \alpha=0.08$)} \\
\\
    15 & -0.08$\pm$0.03 & 0.04$\pm$0.08 & --  & 0.07$\pm$0.08 & --\\
    30 & -0.08$\pm$0.03 & 0.05$\pm$0.08 & --  & 0.10$\pm$0.08 & --\\
    45 & -0.07$\pm$0.02 & 0.09$\pm$0.07 & --  & 0.12$\pm$0.09 & --\\
    60 & -0.06$\pm$0.02 & 0.10$\pm$0.08 & --  & 0.16$\pm$0.08 & --\\
    75 & -0.06$\pm$0.03 & 0.13$\pm$0.09 & --  & 0.20$\pm$0.09 & 0.11$\pm$0.07 \\
\hline
\multicolumn{6}{c}{(Model F -- $a_*=0, \; L/L_{\rm Edd} = 0.36, \; \alpha=0.03$, 1-loop)} \\
\\
    15 & -0.13$\pm$0.07 & -0.07$\pm$0.14 & --  & -0.03$\pm$0.15 & --\\
    30 & -0.11$\pm$0.07 & -0.05$\pm$0.13 & --  & -0.01$\pm$0.14 & --\\
    45 & -0.10$\pm$0.06 & -0.02$\pm$0.13 & --  & 0.02$\pm$0.14 & --\\
    60 & -0.09$\pm$0.06 & -0.01$\pm$0.13 & --  & 0.06$\pm$0.13 & --\\
    75 & -0.09$\pm$0.06 & 0.01$\pm$0.13 & --  & 0.10$\pm$0.14 & 0.09$\pm$0.13 \\
\hline
\end{tabular}

\begin{tablenotes}
\footnotesize
\item Note -- All spectra were fit with BHSPEC disc spectra, where $\alpha_{\rm fit}=0.1$.  The quoted uncertainties correspond to the systematic GRMHD luminosity profile uncertainties (determined empirically by analysing the last 5 subsequent 1000M time chunks of the GRMHD simulations).  The spectral fitting statistical uncertainties were negligibly small ($\Delta a_* \sim 0.001$).  This is because the disc parameters that are responsible for most of the uncertainty (mass, distance, inclination) were not allowed to vary during the fit.
\item
\item $\dagger$  These two rightmost columns correspond to a disc model that uses an alternative measure of the GRMHD luminosity (See \S\ref{sec:advection} and Appendix \ref{app:cooling} for details).  The luminosity profiles corresponding to GRMHD2 are shown in Figure \ref{fig:lum}.
\end{tablenotes}

\end{threeparttable}

\end{center}

\label{tab:discspin}

\end{table*}

The fitting is handled through \textsc{xspec}, a spectral fitting software package commonly used by X-ray astronomers \citep{arnaud96}.  For each spectrum, we fit for only two parameters: BH spin, and mass accretion rate.  We fix the mass, distance, and inclination to exactly the values used to generate the simulated observations.  In the cases where a power-law fit via \textsc{simpl} is needed, we fit for two additional power-law parameters (essentially the normalization and the spectral slope of the power-law, which \textsc{simpl} encapsulates as the scattering fraction of soft disc photons and the photon power-law index).  We list in Table \ref{tab:discspin} our spin results from this fitting exercise.  The first three columns correspond to the following:

\begin{description}
\item `NT' -- We perform a \textsc{bhspec} fit on spectra computed from a \citetalias{NT} disc model, with disc parameters corresponding to Table \ref{tab:discpar}.  We use this as a baseline for comparing with our GRMHD fit results.  For the fit, we use an energy range of 0.4--8.0 keV.\\
\item `GRMHD' -- We perform a \textsc{bhspec} fit on the GRMHD disc spectra on the energy range 0.4--8.0 keV.  We employ an 8.0keV upper energy cutoff in the fitted spectrum since we want to exclude the power-law spectral feature.\\
\item `GRMHD+PL' -- In addition to continuum fitting, we also fit for the power-law.  Formally, we fit a \textsc{bhspec}$\otimes$\textsc{simpl} model over the energy range 0.4--50.0$\,$keV (the full range of our mock observation).\\
\end{description}

\begin{figure}
\begin{center}  
\includegraphics[width=0.5\textwidth]{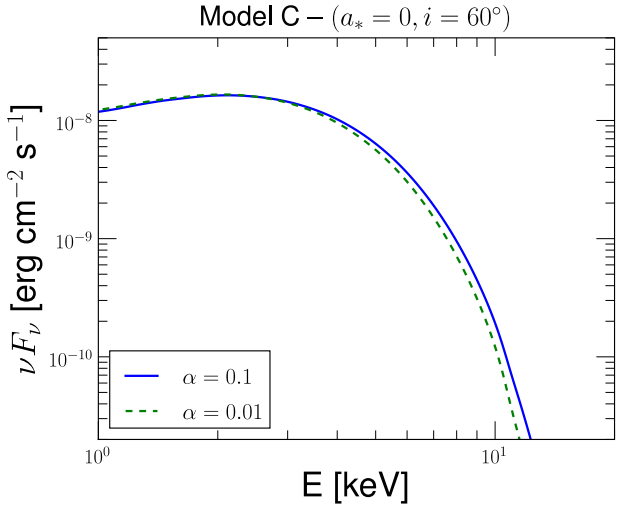} 
\caption{\citetalias{NT} disc spectra for model C ($a_*=0$) computed for two choices of $\alpha$.  The slight shift in the location of the spectral peak causes a systematic error in the recovered BH spin if $\alpha_{\rm fit}$ does not match the instrinsic $\alpha$ of the disc.\label{fig:alphaSpinEffect}}  
\end{center}  
\end{figure}

The primary purpose of the `NT' column is to disentangle another systematic effect that is independent of the extra luminosity of the GRMHD simulations.  The \textsc{bhspec} spectral model is tabulated only for $\alpha_{\rm fit}=0.1$, however in our disc models, only the $a_*=0.7$ disc happens to have a matching $\alpha=0.1$ (see Model B in Table \ref{tab:discpar}).    This mismatch between the fitting model $\alpha_{\rm fit}$ and the input model $\alpha$ leads to a systematic bias in the recovered spin.  The spectral dependence on $\alpha$ arises from the fact that column density scales inversely with $\alpha$, so discs with higher $\alpha$ values tend towards lower optical depths.  A lower optical depth medium has gas that is hotter, which results in harder spectra.  In Figure \ref{fig:alphaSpinEffect}, we illustrate the influence that the choice of $\alpha$ has on the spectum of model C ($a_*=0$).  As expected, we find that the higher $\alpha$ disc has a harder spectrum (Fig. \ref{fig:alphaSpinEffect}).

In the context of spin fitting, if $\alpha_{\rm fit} > \alpha$, we expect the fitted spins to be too low (since for any given spin, the fitting spectrum will be harder than the intrinsic disc spectrum), and vice versa for $\alpha_{\rm fit} < \alpha$.  This effect is seen in the `NT' column of Table \ref{tab:discspin} (which differs from the fitting model only by its choice of $\alpha$).  Models C, D, E, and F all have $\alpha_{\rm fit} > \alpha$, and consequently the recovered spins are too low.  On the other hand, model A has $\alpha_{\rm fit} < \alpha$, yielding recovered spins that are too high.

We interpret the difference between the `NT' column and the `GRMHD' column as the effect of the extra luminosity from the inner regions of the GRMHD discs, which is to systematically increase the recovered spin.  The spin deviation also grows with inclination angle (see the `GRMHD' column in Table \ref{tab:discspin}), since the net effect of Doppler beaming (which becomes more important at higher inclinations) is to enhance the inner disc emission.  We also find that the spin deviation becomes smaller at high spins (this is due to the fact that the ISCO-spin relation becomes steeper at high spins).

In those cases where using the \textsc{simpl} model significantly improves the fit (i.e. when the fit $\chi^2$ improved by more than a factor of 2, upon inclusion of a power-law fit component), the modelling of the power-law also leads to better agreement between fitted and input BH spin (compare `GRMHD' and `GRMHD+PL' columns in Table \ref{tab:discspin}).  By fitting a power-law component, some of the hot photons get associated with the power-law, which lowers the fitted spin (since the remaining disc spectrum looks softer after removal of the hard power-law photons).  However, we note that the GRMHD luminosity excess extends even beyond the ISCO (see Fig. \ref{fig:lum}).  This explains why even fitting for the power-law does not completely eliminate the upwards spin bias (i.e. the `GRMHD+PL' columns still yield spins that are higher than the `NT' column in Table \ref{tab:discspin}); although some of the excess luminosity is bound up in the nonthermal plunging region (which can be excluded by means of power-law fitting), there is still some residual thermal disc luminosity excess, that acts to bias the GRMHD spin upwards.  Overall, the effect of the extra plunging region light results in $\Delta a_*\sim 0.1,\,0.06,\,0.03$ corresponding to spins $a_*=0,\,0.7,\,0.9$ respectively (estimated by comparing the `NT' column with the `GRMHD' column in Table \ref{tab:discspin} for the $\sim$30 per cent Eddington discs).  

  These systematic spin errors should be thought of as an upper bound since the discs that we consider correspond to fairly high accretion rates (with $L/L_{\rm Edd} > 0.3$).  Previous work (\citealt{paczynski00,shafee08a}) suggests that thinner discs (corresponding to lower accretion rates) better match the \citetalias{NT} model.  Our results in Table \ref{tab:discspin} also support this claim; in comparing the spin fits from model C (thin disc) with models D and E (thick discs), we find that model C agrees best to the NT disc (i.e. that model C has the smallest difference between the `NT' and `GRMHD' columns).  In addition, model C also exhibits the least deviation from a thermal spectrum at high energies (compare the $>20\,{\rm keV}$ emission from model C with those of D and E in Fig. \ref{fig:spectraPlot}).

We did not simulate lower accretion rate discs due to the rapidly increasing computational cost associated with such simulations.  Typically, continuum fitting is not applied to systems where the accretion rate exceeds 30 per cent Eddington since beyond this critical point, the continuum fitting method is no longer robust to variations in the disc accretion rate (see \citealt{mcclintock06,steiner09b}).  One possible explanation is that disc self-shadowing becomes important beyond this critical accretion rate threshold \citep{li10}.   

To put into perspective how significant our quoted spin deviations are, we compare our results to the spin uncertainties from actual continuum fitting exercises in the literature (See Table \ref{tab:literatureSpinFits}).  Since the current observational uncertainties are significantly larger than the deviations that we find in this work, we conclude that the extra light from the inner and plunging regions of the disc do not limit our ability to make accurate spin estimates through the continuum fitting method.  The current limiting factors in measuring BH spin are how accurately one can determine the distance, mass, and inclination of black hole binary systems.

\begin{table*}
\caption{Spin measurements yielded by continuum fitting.\label{tab:literatureSpinFits}}

\begin{center}

\begin{threeparttable}

\begin{tabular}{lccccr}
\hline
Black hole & $a_*$ & $i(^\circ)$ & $L/L_{\rm Edd}$ & $\alpha_{\rm fit}$ & Reference \\
\hline
A0620-00            & 0.12 $\pm$ 0.19          & $51.0\pm0.9$     & 0.11              & 0.01+0.1\tnote{a}    & \citet{gou10} \\

H1743-322\tnote{c}     & $ 0.2\pm0.3$\tnote{c}            & $75\pm3$         & 0.03-0.3          & 0.01+0.1\tnote{a}    & \citet{steiner12}\\

LMC X-3             & $<0.3$                   & $67\pm2$         & 0.1-0.7           & 0.01, 0.1\tnote{b}   & \citet{davisdone06}\\

XTE J1550-564       & $0.34^{+0.2}_{-0.28}$    & $74.7\pm3.8$     & 0.05-0.30         & 0.1               & \citet{steiner11}\\

GRO J1655-40\tnote{d}  & $0.70\pm0.1$           & $70.2\pm1.2$     & 0.04-0.1          & 0.1               & \citet{shafee06}\\

4U 1543-47\tnote{d}    & $0.80\pm0.1$           & $20.7\pm1.5$     & 0.06-0.1          & 0.1               & \citet{shafee06}\\

M33 X-7             & $0.84\pm0.05$            & $74.6\pm1.0$     & 0.07-0.11         & 0.01              & \citet{liu08,liu10}\\

LMC X-1             & $0.92^{+0.05}_{-0.07}$   & $36.4\pm2.0$     & 0.15-0.17         & 0.01+0.1\tnote{a}    & \citet{gou09}\\

GRS 1915+105\tnote{d}  & $>0.95$          &  61.5-68.6       & 0.2-0.3           &  0.01, 0.1\tnote{b}  & \citet{mcclintock06}\\

Cygnus X-1          & $>0.95$                  & $27.1\pm0.8$     & 0.018-0.026       & 0.01+0.1\tnote{a}    & \citet{gou11}\\

\hline
\end{tabular}
\begin{tablenotes}
\item Note -- The spin uncertainties correspond to the 68 per cent (1$\sigma$) level of confidence, whereas the inequalities are to the 3$\sigma$ level.
\item
\item [a] The spin errors are marginalized over both choices of $\alpha$
\item [b] The spin constraint covers both choices of $\alpha$
\item [c] No reliable mass estimate is available for this source
\item [d] The quoted spin errors have not been rigorously computed, and have been arbitrarily doubled from the published estimates  since these are among the first systems for which continuum fitting was applied).

\end{tablenotes}

\end{threeparttable}

\end{center}

\end{table*}


\section{Discussion}\label{sec:discussion}

\begin{figure}
\begin{center}  
\includegraphics[width=0.5\textwidth]{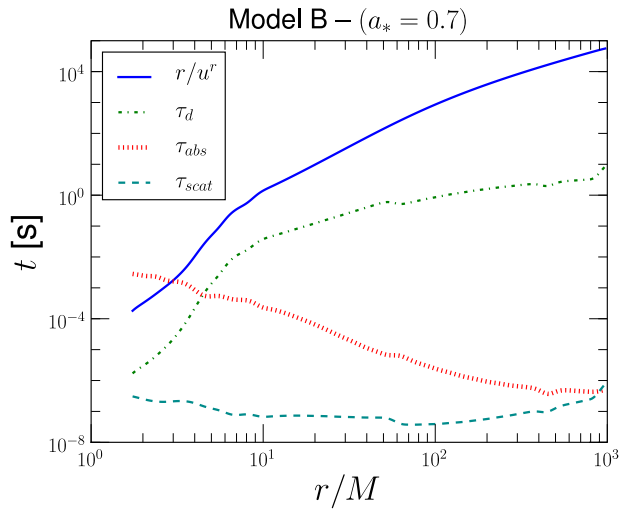} 
\caption{Several time-scales of interest are plotted here for model B ($a_*=0.7$).  We show the accretion time-scale (solid blue), the photon diffusion time-scale (dash-dotted green), the photon absorption time-scale (dotted red), and the photon scattering time-scale (dashed dark cyan).  The absorption and scattering time-scales are computed via $t=1/(\rho \kappa c)$, using midplane gas densities ($\rho$) and opacities ($\kappa$).  The diffusion time-scale is computed via $t_{\rm d} = \tau_{\rm scat}^2 t_{\rm scat}$ where $\tau_{\rm scat}$ is the scattering optical depth to the disc midplane (i.e. we expect $\tau_{\rm scat}^2$ scatterings to occur before a photon can diffuse out).  Note that close to the black hole, the accretion time-scale becomes shorter than the photon absorption time-scale, which indicates that the fluid close to the black hole has insufficient time to reach thermal equilibrium.\label{fig:time-scales}}  
\end{center}  
\end{figure}

The disc model that we adopt in this work is not completely self-consistent.  The annuli that we compute using \textsc{tlusty} are assumed to have reached their equilibrium configuration.  However, close to the black hole there is simply insufficient time for the fluid to reach equilibrium (see Fig. \ref{fig:time-scales} for a plot of the various time-scales).  In addition, the GRMHD simulations show that the pressure in the innermost regions of the disc is magnetically dominated, yet in the \textsc{tlusty} annuli calculations, we completely ignore magnetic pressure support.  Previous work by \citet{davis09} suggests that including magnetic support may lead to a slight hardening of the annuli spectrum.  Another problem for the innermost annuli is that they become extremely hot ($T_{\rm gas} > 10^8 - 10^9$ K), which may invalidate \textsc{tlusty}'s handling of Comptonization (i.e. an angle averaged Kompaneets treatment, which is only valid for nonrelativistic electrons, see \citealt{hubeny01}).  Finally, due to the increasingly strong light bending effects near the black hole, the process of disc self-irradiation (i.e. where one part of the disc shines on another part) may also become important, especially for the innermost annuli \citep{li05}.  The net effect of this disc self-irradiation would be to further heat up the annuli, causing additional spectral hardening.  We have ignored disc self-irradiaton in our spectral model since this reprocessing of light cannot be easily handled by our grid of precomputed annuli.

These issues primarily affect annuli that are very close to the black hole, which radiate in the hard X-rays ($>20\,$keV).   We do not expect these inconsistent annuli to affect our spin fitting results in Table \ref{tab:discspin} since the recovered spin is mainly constrained by photons near the peak of the spectrum (located at a few keV in Figure \ref{fig:spectraPlot}).  The inconsistent annuli will however affect the high energy part of the spectrum, which is why we caution against reading too deeply into the quantitative details of the power laws.

\subsection{Other signatures of the plunging region}

Observations of disc variability could potentially be used to discriminate between our plunging region model and other coronal models.  Assuming that the variability is caused by Doppler beaming of hotspots moving about in the disc flow,  we associate the hotspot orbital time-scale with the variability time-scale.  Since the orbital time decreases monotonically as one approaches the horizon, the variability from the inner plunging region of the disc ought to be more rapid than that occurring farther out in disc.  Hence, our plunging region model predicts that the hard X-ray light (originating from the innermost regions of the disc) would have faster variability than the soft X-ray light (produced farther out in the thermal disc).  Furthermore, in the power-density spectrum in the hard X-ray bands, we expect most of the variability power to occur above the ISCO orbital frequency (since we associate this hard emission with the plunging region).

Polarization observations could also serve as a means to distinguish between other coronal models and a plunging region model. Generally, one expects evolution of the polarized fraction and net polarization angle as the observed photon energy is varied. This is due to higher energy photons (which are typically emitted closer to the black hole) feeling stronger relativistic effects, leading to a stronger shift in their polarization vectors \citep{schnittman09,schnittman10}. The planar geometry of the plunging region may lead to a polarization signature that differs significantly from coronal models. If the high-energy power law originates from the plunging region, we expect that its polarization properties will vary continuously from the thermal component to the non-thermal power law. This contrasts with some alternative coronal geometries (e.g. spherical models) that provide abrupt transitions in the polarization properties from one component to the next \citep{schnittman10}.

In the plunging region model, even at energies where the power law dominates, the variation of the polarization fraction and angle with energy could be consistent with thermal state polarization models that assume a thin disc geometry.  The signal might then be similar to the models of \citet{li09,schnittman09}, who calculate the polarization due to direct emission from a NT model, assuming a (modified) blackbody spectrum and intrinsic disc polarization appropriate for a scattering dominated atmosphere.  The models also include the effect of scattered emission from light that is strongly bent by the relativistic spacetime of the black hole and reintersects the disc surface (i.e. returning radiation).  In these calculations, matter in the plunging region is allowed to scatter returning radiation, but does not provide direct emission.  Since the intrinsic emission and returning radiation have different polarization signals, a similar calculation that includes the direct (power law) emission from the plunging region is required to confirm the above speculation and discriminate between coronal models. The impending launch of NASA’s Gravity and Extreme Magnetism Small Explorer (\citealt{swank10}; GEMS) mission promises observational constraints that may be capable of discriminating between such models.

\subsection{How does our choice of cooling function influence the results?}\label{sec:advection}

A criticism that might be levied against the GRMHD simulations is that our choice of cooling function (as defined in Eq. \ref{eq:cooling}) is arbitrary.  We have interpreted this cooling function as the rate of radiative cooling, despite the fact that the GRMHD simulations do not inherently model radiation physics.  

We would like to test how our choice of cooling function impacts the results of \S\ref{sec:spinResults}. Rather than rerunning the GRMHD simulations with another prescription for cooling (i.e. by changing the functional form of Eq. \ref{eq:cooling}), we use the simulation's dissipative heating profile as another means to get a cooling rate.  Instead of simply assuming that dissipation equals cooling locally, we consider the physics of energy advection (i.e. that the fluid releases its heat downstream).  To generate a disc model that is more self-consistent with the \textsc{TLUSTY} annuli, we use the \textsc{TLUSTY} vertical structure when computing the rate of energy advection (see Appendix \ref{app:cooling} for a thorough exposition on this method of computing the new GRMHD luminosity profile).  The new dissipation-based cooling function (which we label as GRMHD2) releases slightly more luminosity inside the plunging region (compare the solid and dashed lines in Figure \ref{fig:lum}).

We repeat the same exercise as in \S\ref{sec:spinResults} with this new GRMHD2 derived luminosity profile.  The spectra corresponding to GRMHD2 are shown as the dashed lines in Figure \ref{fig:spectraPlot}.  Without fitting for the power laws, we find that the recovered spins are slightly higher owing to the extra plunging region luminosity (compare the `GRMHD' and `GRMHD2' columns in Table \ref{tab:discspin}).  However, the more realistic fit (i.e. including the power-law component) takes care of this extra light, yielding comparable spin estimates to before.  The enhanced power laws are purely due to the extra plunging region luminosity (compare solid and dashed lines in Figures \ref{fig:lum} and \ref{fig:spectraPlot}).  The worst cases for the $30$ per cent Eddington discs now have $\Delta a_*$ $\sim$0.15, 0.07, 0.03, for $a_*=0,\;0.7,\;0.9$ respectively.  Given these results, we still conclude that the extra luminosity from the GRMHD inner disc does not limit our ability to measure BH spin through the continuum fitting method (the dominant source of uncertainty is still the observational uncertainties in distance, mass, and inclination).
 
As in \citetalias{kulkarni11}, we estimate that $\Delta a_* \leq 0.15$, which is lower than the estimate of Noble et al. (2011), who infer $\Delta a \sim 0.2-0.3$ for a Schwarzschild black hole.  Due to the reduced sensitivity of the continuum fitting method at low spins, this amounts to a somewhat modest discrepancy, which we attribute to various differences in the setup and intitial conditions of the GRMHD simulations used by the different groups.

\subsection{What is the impact of the initial magnetic field topology?}

It can be argued that our choice of four weak poloidal magnetic loops is arbitrary, and that the most natural choice is either a singly looped purely toroidal or purely poloidal field \citep{igumenshchev03}.  Our choice is motivated by the belief that nature does not begin the accretion process with an ordered field spanning large spatial scales;  starting from a weak and disordered field, we believe the physics of the magneto-rotational instability (MRI) naturally selects the correct magnetic configuration, forgetting about the initial conditions.  Ideally, we would like to start the disc with an infinite number of loops to model this tangled geometry, however owing to resolution limitations we settle with four loops.

To determine how much influence the initial field geometry could have on disc spectra, we compare the results of Model C (4-loop) and Model F (1-loop).  In general, we find that the 1-loop case releases significantly more luminosity inside the plunging region.  The large-scale magnetic fluxes from the 1-loop geometry produce more stress and dissipation than its 4-loop counterpart.  However, most of this energy is released at large polar angles (i.e. far from the disc midplane where the gas is unbound), and for the purposes of modelling the thermal disc emission, we choose to ignore this unbound gas contribution to the luminosity.  

Making this distinction between bound and unbound gas, we find that the luminosity profile corresponding to the bound gas in the 1-loop case closely matches that of the 4-loop case.  This leads to very similar disc spectra (compare the spin fit results from Models C and F in Table \ref{tab:discspin} and spectra in Figure \ref{fig:spectraPlot}).  We conclude that in the context of the thermal continuum for thin discs, the resultant spectrum is largely independent of the initial magnetic field geometry.  This is in contrast to the case of magnetically arrested accretion flows, whose final accretion state and jet power depends crucially on both the strength and geometry of its seed magnetic field \citep{igumenshchev08,igumenshchev09,tchekhovskoy11,mckinney12}.  For thin-discs, the primary differences arising from the choice of 1-loop and 4-loop configurations are highlighted in \S8 and \S9 of \citet{penna10}.   For a complete discussion of this topic, see also: \citetalias{kulkarni11}; \citet{noble10,noble11}; and \citet{hawley11}.

\subsection{Convergence of GRMHD simulations}

Recent work by \citet{hawley11} suggests that the growth and saturation of the MRI in disc simulations is highly dependent on their vertical and azimuthal resolution.  These authors propose the following two convergence criteria for the MRI to be well resolved: the vertical criterion is given by $Q_z=\lambda_{\rm MRI}/(\Delta z) \gtrsim 10$, and the azimuthal criterion $Q_y=\lambda_{\rm MRI}/(\Delta y) \gtrsim 5$, where $\lambda_{\rm MRI}$ is the MRI length scale in the direction that Q is being measured.

To compare with \citet{hawley11}, we adopt their prescription for $\lambda_{\rm MRI}=2\pi H \beta^{-1/2}|B_z|/|B|$ in the evaluation of $Q_z$ and $\lambda_{\rm MRI}=2\pi H \beta_{y}^{-1/2}$ in $Q_y$, where $H$ is the disc scale height, plasma $\beta=P_{\rm gas}/P_{\rm mag}$, and plasma $\beta_y$ is computed with only the toroidal field component.  For the 4-loop standard  resolution runs\footnote{This includes models A-D in Table \ref{tab:discpar}.  Model E, which has double the resolution, will have double the $Q_z$ and $Q_y$ values.}, we find $Q_z = 6 - 13$ at the disc midplane and $Q_y = 2 - 3$.  

As an alternative model-free means to estimate the MRI length scale, we have also performed a correlation length analysis similar to the one presented in \S3.8 of \citet{mckinney12}.  From the spatial autocorrelation function for gas density, we find $Q_z \sim 9-11$ and $Q_y \sim 5-7$.  Azimuthal shear elongation of the MRI unstable zone is the reason why the correlation length based $Q_y$ is larger than the \citet{hawley11} estimate for $Q_y$.

Since our value for $Q_y$ is low, this implies that the simulations are only marginally resolving the azimuthal disc substructure (and hence MRI turbulence).  A consequence of this low resolution is that the strength of the MRI turbulence (as measured by the disc viscosity $\alpha$) does not appear to converge with respect to resolution.  Comparing models D and E in Table \ref{tab:discpar}, we find that $\alpha$ has doubled upon doubling both $\phi$ and $\theta$ resolutions.  However, recent work by \citet{sorathia12} suggests that $\alpha$ is a poor metric for gauging simulation convergence since large variations in $\alpha$ are observed even in the cases of abundant resolution.

For the purposes of spectral modelling we ultimately only care about the simulation luminosity profile.  Despite the low azimuthal resolution of the \citet{penna10} simulations, the luminosity profile remains fairly insensitive to  changes in $\phi$-resolution.  The convergence tests performed by \citet{penna10}  show that the luminosity profile changes only by ~10\%, even after a 4-fold increase in $\phi$-resolution.  This robustness in luminosity profile is also borne out when comparing the spectra of Model D and Model E (see Figure \ref{fig:spectraPlot} -- these two runs only differ in grid resolution).

\subsection{Equation of state}

Another inconsistency in the GRMHD simulations is that we adopt an ultrarelativistic gas equation of state (with $\Gamma=4/3$), motivated by the idea that the disc energy budget is dominated by photons (a $\Gamma=4/3$ fluid).  However the \textsc{tlusty} annuli calculations suggest that the combined photon and gas entity acts more like a $\Gamma=5/3$ fluid, especially in the gas pressure dominated plunging region. Despite the difference in $\Gamma$, \citet{noble09,noble10,noble11} have also performed GRMHD simulations of thin discs with $\Gamma=5/3$, and they find similar results to the $\Gamma=4/3$ discs of \citet{shafee08b,penna10}.

\begin{figure}
\begin{center}  
\includegraphics[width=0.5\textwidth]{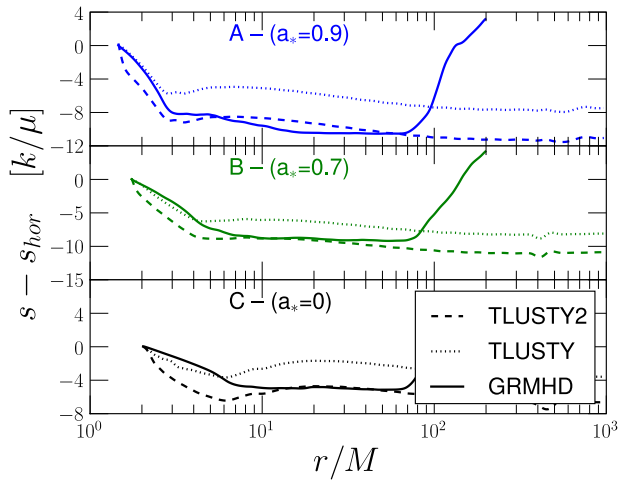} 
\caption{GRMHD vs. \textsc{tlusty} vertically averaged specific gas entropy profiles (relative to the entropy at the horizon $s_{hor}$). The dimensionless entropy is computed using an ideal gas equation of state, where $s=\left(\frac{k}{\mu}\right)\cdot\frac{1}{\Gamma-1}\ln(p_{\rm gas}/\rho^\Gamma)$. The TLUSTY lines (dotted) denote the gas entropy as computed from the TLUSTY annuli vertical structure.  The TLUSTY2 lines (dashed) represent annuli that comprise the GRMHD2 disc model (described in \S\ref{sec:advection} and Appendix \ref{app:cooling}).\label{fig:s}}  
\end{center}  
\end{figure}

In figure \ref{fig:s}, we examine the relative impact that this equation of state inconsistency has on the disc's entropy profile.  A fully self-consistent model would have identical GRMHD and \textsc{tlusty} gas entropy profiles.  It appears that well outside the ISCO, the assumption that the gas cools towards a fixed entropy is supported by the radiative transfer calculations of \textsc{tlusty} (Compare the flat plateaus in Figure \ref{fig:s}).  However Figure \ref{fig:s} shows that the GRMHD (solid) and TLUSTY (dotted) discs reach different constant entropies.  The alternate disc cooling model (dashed) discussed in \S\ref{sec:advection} yields annuli which are more self-consistent with the simulation entropy profile (compare dashed lines and solid lines in Figure \ref{fig:s}, which now track each other fairly well outside the ISCO).\\[0.5in]


\section{Summary}\label{sec:summary}

The primary goal of this work is to determine how important the neglected light from the plunging region is in the context of BH spin measurements.   To answer this question, we rely on GRMHD disc simulations of \citet{penna10}, which capture all but the radiation physics for magnetized flow around a black hole. We seek to convert these dimensionless simulations into a form that can be directly compared with observations, namely the accretion discs' X-ray continuum spectra.

To do this, we apply radiative transfer post-processing on the simulated discs.  This is an advance over previous work on computing GRMHD disc spectra \citepalias[i.e.][]{kulkarni11}, which assumes (modified) blackbody annuli spectra.  We slice the GRMHD disc into many individual annuli, and for each annulus, we apply a 1-dimensional radiative transfer calculation to solve for its emergent spectrum.  We sum up this collection of local spectra by means of ray tracing to get the full disc spectrum, which we then compare with contemporary disc models used to measure black hole spin.  The following are the key results from this work:

\begin{enumerate}
\renewcommand{\theenumi}{(\arabic{enumi})}
\item The GRMHD based accretion discs have hotter spectra than the standard \citetalias{NT} discs.  The GRMHD discs produce more luminosity everywhere, and the contrast becomes most apparent inside the ISCO (see Figure \ref{fig:lum}).\\
\item The increased luminosity of the GRMHD discs compared to the classic NT discs induces a modest systematic bias in the derived spins of these GRMHD discs.  For black holes of spin $a_*=0,\,0.7,\,0.9$ the spin deviation is $\Delta a_* \sim 0.15,\,0.07,\,0.03$ in the worst cases (corresponding to inclination angles of 75$^\circ$).  We remark that these errors are well within observational uncertainties (i.e. from not precisely knowing the system's mass, inclination, and distance -- see Table \ref{tab:literatureSpinFits}).\\
\item Without needing to invoke an external corona, the GRMHD discs around spinning black holes exhibit a weak high-energy power-law tail (Fig. \ref{fig:spectraPlot}).  This power-law tail arises from the combined emission of the hot plunging region gas.  The strength of this plunging region power-law increases with the system's inclination angle.\\
\end{enumerate}

In our spectral modelling approach, we have made many simplifications and assumptions (see \S\ref{sec:discussion}), some of which may be incorrect close to the black hole horizon.  Due to these problems, we trust the power laws presented here only to a qualitative level -- we only trust that the plunging region fluid emits at much higher energies than the disc, which adds a nonthermal component to the overall spectrum at high energies.

\section{Acknowledgments}

The authors would like to thank Jack Steiner, Olek S\c{a}dowski, and Lijun Gou for stimulating discussions and helpful suggestions.  We also thank the anonymous referee for their insightful comments and feedback.  We acknowledge support from NSF grant AST-0805832.  The GRMHD simulations were made possible by the NSF through TeraGrid resources provided by NCSA (Abe), LONI (QueenBee), and NICS (Kraken), and by the NASA cluster Pleiades.  TLUSTY spectral models were computed using the Odyssey cluster supported by the FAS Science Division Research Computing Group at Harvard University.


\bibliography{ms}

\begin{thebibliography}{65}
\expandafter\ifx\csname natexlab\endcsname\relax\def\natexlab#1{#1}\fi


\bibitem[{Abramowicz} et~al.(1988)]{abram88}
{Abramowicz} M.~A., {Czerny} B., {Lasota} J.~P., {Szuszkiewicz} E., 1988, \apj, 332, 646

\bibitem[{Abramowicz} et~al.(1997)]{abram97}
{Abramowicz} M.~A., {Lanza} A., {Percival} M.~J., 1997, \apj, 479, 179

\bibitem[{Abramowicz} \& {Klu\'{z}niak}(2001)]{abram01}
{Abramowicz} M.~A., {Klu\'{z}niak} W. 2001, \aap, 374:L19-20

\bibitem[{Afshordi} \& {Paczy\'nsky}(2003)]{afshordi03}
{Afshordi} N., {Paczy\'nski} B., 2003, \apj, 592, 354

\bibitem[{Arnaud}(1996)]{arnaud96}
{Arnaud}, K.~A. 1996, in {Jacoby G.~H., Barnes J., eds}, {Astronomical Society of the Pacific Conference Series, Vol. 101, Astronomical Data Analysis Software and Systems V, San Francisco: ASP}, 17

\bibitem[{Beckwith} et~al.(2008a)]{beckwith08}
{Beckwith} K., {Hawley} J.~F., {Krolik} J.~H., 2008a \mnras, 390, 21

\bibitem[{Beckwith} et~al.(2008b)]{beckwith08b}
{Beckwith} K., {Hawley} J.~F., {Krolik} J.~H., 2008b \apj, 678, 1180

\bibitem[{Blandford} \& {Znajek}(1977)]{blandford77}
{Blandford} R.~D., {Znajek} R.~L. 1977, \mnras, 179, 433

\bibitem[{Bolton}(1972)]{bolton72}
{Bolton} C.~T., 1972, Nat, 240, 124

\bibitem[{Cantrell} et~al.(2010)]{cantrell10}
{Cantrell} A.~G. et~al., 2010, \apj, 710, 1127

\bibitem[{Davis} et~al.(2005)]{davis05}
{Davis} S.~W., {Blaes} O.~M., {Hubeny} I., {Turner} N.~J., 2005, \apj, 621, 372

\bibitem[{Davis} \& {Hubeny}(2006a)]{davishubeny06}
{Davis} S.~W., {Hubeny} I. 2006a, \apjs, 164, 530

\bibitem[{Davis} et~al.(2006b)]{davisdone06}
{Davis} S.~W., {Done} C., {Blaes} O.~M., 2006b, \apj, 647, 525

\bibitem[{Davis} et~al.(2009)]{davis09}
{Davis} S.~W., {Blaes} O.~M., {Hirose} S., {Krolik} J.~H., 2009, \apj, 703, 569

\bibitem[{De Villiers} et~al.(2003)]{deVilliers03}
{De Villiers} J., {Hawley} J.~F., {Krolik} J.~H., 2003, \apj, 599, 1238

\bibitem[{Doeleman} et~al.(2009)]{doeleman09}
{Doeleman} S. et~al., 2009, Astro2010: The Astronomy and Astrophysics Decadal Survey, Science White Papers, No., 68

\bibitem[{Dove} et~al.(1997)]{dove97}
{Dove} J.~B., {Wilms} J., {Begelman} M.~C. 1997, \apj, 487, 747

\bibitem[{Fabian} et~al.(1989)]{fabian89}
{Fabian}, A.~C., {Rees}, M.~J., {Stella}, L., {White}, N.~E., 1989, \mnras, 238, 729

\bibitem[{Gammie} et~al.(2003)]{gammie03}
{Gammie} C. F., {McKinney} J. C., {T\'oth} G., 2003, \apj, 589, 444

\bibitem[{Gou} et~al.(2009)]{gou09}
{Gou} L. et~al., 2009, \apj, 701, 1076

\bibitem[{Gou} et~al.(2010)]{gou10}
{Gou} L., {McClintock} J. E., {Steiner} J. F., {Narayan} R., {Cantrell} A. G., {Bailyn} C. D., {Orosz} J. A., 2010, \apj, 718, L122

\bibitem[{Gou} et~al.(2011)]{gou11}
{Gou} L. et~al., 2011, \apj, 742, 85

\bibitem[{Graham} et~al.(2011)]{graham11}
{Graham} A.~W., {Onken} C.~A., Athanassoula E., Combes F., 2011, \mnras, 412, 2211

\bibitem[{G\"ultekin} et~al.(2009)]{gultekin09}
{G\"ultekin} et~al., 2009, \apj, 698, 198

\bibitem[{Haardt} \& {Maraschi}(1991)]{haardt91}
{Haardt} F., {Maraschi} L., 1991, \apj, 380, L51

\bibitem[{Hawley}, {Guan}, \& {Krolik}(2011)]{hawley11}
{Hawley} J.~F., {Guan} X., {Krolik} J.~H., 2011, \apj, 738, 84

\bibitem[{Hubeny} \& {Lanz}(1995)]{hubeny95}
{Hubeny} I., {Lanz} T. 1995, \apj, 439, 875

\bibitem[{Hubeny} et~al.(2001)]{hubeny01}
{Hubeny} I., {Blaes} O., {Krolik} J. H., Agol E., 2001, \apj, 559, 680

\bibitem[{Igumenshchev}(2008)]{igumenshchev08}
{Igumenshchev} I.~V., 2008, \apj, 677, 317

\bibitem[{Igumenshchev}(2009)]{igumenshchev09}
{Igumenshchev} I.~V., 2009, \apj, 702, 72

\bibitem[{Igumenshchev}, {Narayan}, \& {Abramowicz}(2003)]{igumenshchev03}
{Igumenshchev} I.~V., {Narayan} R., {Abramowicz} M.~A., 2003, \apj, 592, 1042

\bibitem[{Jahoda} et~al.(2006)]{jahoda06}
{Jahoda} K., {Markwardt} C.~B., {Radeva} Y., {Rots} A.~H., {Stark} M.~J., {Swank} J.~H., {Strohmayer} T.~E., {Zhang} W., 2006, \apjs, 163, 401

\bibitem[{Kawaguchi} et~al.(2000)]{kawaguchi00}
{Kawaguchi} T., {Shimura} T., {Mineshige} S., 2000, NewAR, 44, 443

\bibitem[{Kulkarni} et~al.(2011)]{kulkarni11}
{Kulkarni} A.~K. et~al., 2011, \mnras, 414, 1183

\bibitem[Li et~al.(2010)]{li10}
{Li} G.-X., {Yuan} Y.-F., {Cao} X., 2010, \apj, 715, 623

\bibitem[Li, Narayan, \& McClintock(2009)]{li09}
{Li} L.-X., {Narayan} R., {McClintock} J.~E., 2009, \apj, 691, 847

\bibitem[{Li} et~al.(2005)]{li05}
{Li} L.-X., {Zimmerman} E. R., {Narayan} R., {McClintock} J. E., 2005, \apjs, 157, 335

\bibitem[{Liu} et~al.(2002)]{liu02}
{Liu} B.~F., {Mineshige} S., {Shibata} K., 2002, \apj, 572, L173

\bibitem[{Liu} et~al.(2008)]{liu08}
{Liu} J., {McClintock} J.~E., {Narayan} R., {Davis} S.~W., {Orosz} J.~A, 2008, \apj, 679, L37

\bibitem[{Liu} et~al.(2010)]{liu10}
{Liu} J., {McClintock} J.~E., {Narayan} R., {Davis} S.~W., {Orosz} J.~A, 2010, \apj, 719 L109

\bibitem[{McClintock} \& {Remillard}(2006)]{mcclintockremillard06}
{McClintock} J.~E., {Remillard} R.~A., 2006, in {Lewin W.~H.~G.}, {van der Klis M.}, eds, {Compact Stellar X-ray Sources, Cambridge Univ. Press, Cambridge}, p. 157

\bibitem[{McClintock} et~al.(2001)]{mcclintock01}
{McClintock} J.~E. et~al., 2001, \apj, 555, 477

\bibitem[{McClintock} et~al.(2006)]{mcclintock06}
{McClintock} J.~E.,  {Shafee} R., Narayan R., Remillard R.~A., Davis S.~W., Li L.-X., 2006, \apj, 652, 518

\bibitem[{McClintock} et~al.(2011)]{mcclintock11}
{McClintock} J.~E. et~al., 2011, Classical and Quantum Gravity, 28, 114009

\bibitem[{MacFadyen} \& {Woolsley}(1999)]{macfadyen99}
{MacFadyen} A.~I., {Woolsley}, S.~E., 1999, \apj, 524, 262

\bibitem[{McKinney}(2006)]{mckinney06}
{McKinney} J.~C., 2006, \mnras, 368, 1561

\bibitem[{McKinney} \& {Blandford}(2009)]{mckinney09}
{McKinney} J.~C., {Blandford} R. D., 2009, \mnras, 394, L126

\bibitem[{McKinney}, {Tchekhovskoy}, \& {Blandford}(2012)]{mckinney12}
{McKinney} J.~C., {Tchekhovskoy} A., {Blandford} R. D., 2012, \emph{arXiv:1201.4163}

\bibitem[{Miller}(2007)]{miller07}
{Miller} J.~M., 2007, ARA\&A, 45, 441

\bibitem[{Mitsuda} et~al.(1984)]{mitsuda84}
{Mitsuda} K. et~al., 1984, \pasj, 36, 741

\bibitem[{Miyamoto} et~al.(1991)]{miyamoto91}
{Miyamoto} S., {Kimura} K., {Kitamoto} S., {Dotani} T., {Ebisawa} K., 1991, \apj, 383, 784

\bibitem[{Noble} et~al.(2009)]{noble09}
{Noble} S.~C., {Krolik} J.~H., {Hawley} J.~F., 2009, \apj, 692, 411

\bibitem[{Noble} et~al.(2010)]{noble10}
{Noble} S.~C., {Krolik} J.~H., {Hawley} J.~F., 2010, \apj, 711, 959

\bibitem[{Noble} et~al.(2011)]{noble11}
{Noble} S.~C., {Krolik} J.~H., {Snittman} J.~D., {Hawley} J.~F., 2011, \apj, 743, 115

\bibitem[{Novikov} \& {Thorne}(1973)]{NT}
{Novikov} I.~D., {Thorne} K.~S., 1973, in {Dewitt} C., {Dewitt} B.~S., eds, {Black Holes (Les Astres Occulus), Gordon and Breach, Paris}, p.343

\bibitem[{Orosz} et~al.(2007)]{orosz07}
{Orosz} J.~A. et~al., 2007, Nat, 449, 872

\bibitem[{Orosz} et~al.(2009)]{orosz09}
{Orosz} J.~A. et~al., 2009, \apj, 697, 573

\bibitem[{Orosz} et~al.(2011a)]{orosz11a}
{Orosz} J.~A., {Steiner} J.~F., {McClintock} J.~E., {Torres} M.~A.~P., {Remillard} R.~A., {Bailyn} C.~D., {Miller} J.~D., 2011, \apj, 730, 75

\bibitem[{Orosz} et~al.(2011b)]{orosz11b}
{Orosz} J.~A., {McClintock} J.~E., {Aufdenberg} J.~P., {Remillard} R.~A., {Reid} M.~J., {Narayan} R., {Gou} L., 2011, \apj, 742, 84

\bibitem[{Paczy\'nski}(2000)]{paczynski00}
{Paczy\'nski} B., 2000, astro-ph/0004129

\bibitem[{Page} \& {Thorne}(1974)]{pagethorne74}
{Page} D.~N., {Thorne} I.~D. 1974, \apj, 191, 499

\bibitem[{Penna} et~al.(2010)]{penna10}
{Penna} R.~F., {McKinney} J.~C., {Narayan} R., {Tchekovskoy} A., {Shafee} R., {McClintock} J.~E., 2010, \mnras, 408, 752

\bibitem[{Reid} et~al.(2011)]{reid11}
{Reid} M.~J., {McClintock} J.~E., {Narayan} R., {Gou} L., {Remillard} R.~A., {Orosz} J.~A., 2011, \apj, 742, 83

\bibitem[{Reis} et~al.(2010)]{reis10}
{Reis} R. C., {Fabian}, A. C., {Miller}, J. M., 2010, \mnras, 402, 836

\bibitem[{Remillard} \& {McClintock}(2006)]{remillard06}
{Remillard} R.~A., {McClintock}, J.~E., 2006, ARA\&A. 44, 49

\bibitem[{Riffert} \& {Herold}(1995)]{riffert95}
{Riffert} H., {Herold} H., 1995 \apj, 450, 508

\bibitem[{Rybicki} \& {Lightman}(1979)]{rybicki79}
{Rybicki} G.~B., {Lightman} A.~P., 1979, {Radiative Processes in Astrophysics, John Wiley \& Sons, New York}, p39

\bibitem[{S\c{a}dowski}(2009)]{sadowski09}
{S\c{a}dowski} A., 2009, \apjs, 183, 171

\bibitem[{S\c{a}dowski} et~al.(2011)]{sadowski11}
{S\c{a}dowski} A., {Abramowicz} M., {Bursa} M., {Klu\'{z}niak} W., {Lasota} J.~P., R\'{o}$\dot{\rm z}$a\'{n}ska A., 2011, \aap, 527A, 17

\bibitem[{Scherbakov} \& {Huang}(2011)]{scherbakov11}
{Scherbakov} R.~V., {Huang} L., 2011, \mnras, 410, 1052

\bibitem[{Schnittman} \& {Krolik}(2009)]{schnittman09}
{Schnittman} J.~D., {Krolik} J.~H.\, 2009, \apj, 701, 1175

\bibitem[{Schnittman} \& {Krolik}(2010)]{schnittman10}
{Schnittman} J.~D., {Krolik} J.~H.\, 2010, \apj, 712, 908

\bibitem[{Shafee} et~al.(2006)]{shafee06}
{Shafee} R., {McClintock} J.~E., {Narayan} R., {Davis} S.~W., {Li} L., {Remillard} R.~A., 2006, \apj, 636, L113

\bibitem[{Shafee} et~al.(2008a)]{shafee08a}
{Shafee} R., {Narayan} R., {McClintock} J.~E., 2008a, \apj, 676, 549

\bibitem[{Shafee} et~al.(2008b)]{shafee08b}
{Shafee} R., {McKinney} J.~C., {Narayan} R., {Tchekovskoy} A., {Gammie} C.~F., {McClintock} J.~E., 2008b, \apj, 687, L25

\bibitem[{Shakura} \& {Sunyaev}(1973)]{shakura73}
{Shakura} N.~I., {Sunyaev} R.~A., 1973, \aap, 24, 337

\bibitem[{Shapiro}, {Lightman}, \& {Eardley}(1976)]{shapiro76}
{Shapiro} S. L., {Lightman} A. P., {Eardley} D. M., 1976, \apj, 204, 187

\bibitem[{Shapiro} \& {Teukolsky}(1983)]{shapiro83}
{Shapiro} S.~L., {Teukolsky} S.~A., 1983, {Black Holes,White Dwarfs, and Neutron Stars: The Physics of Compact
Objects, Wiley, New York}, p. 362.

\bibitem[{Sorathia}, et~al.(2012)]{sorathia12}
{Sorathia} K.~A., {Reynolds} C.~S., {Stone} J.~M., {Beckwith} K., 2012, \apj, 749, 189

\bibitem[{Steiner} et~al.(2009a)]{steiner09a}
{Steiner} J.~F., {Narayan} R., {McClintock} J.~E., {Ebisawa} K., 2009a, PASP, 121, 1279

\bibitem[{Steiner} et~al.(2009b)]{steiner09b}
{Steiner} J.~F., {McClintock} J.~E., {Remillard} R.~A., {Narayan} R., {Gou} L., 2009b, \apj, 701, L83

\bibitem[{Steiner} et~al.(2011)]{steiner11}
{Steiner} J.~F. et~al., 2011, \mnras, 416, 941

\bibitem[{Steiner} et~al.(2012)]{steiner12}
{Steiner} J.~F., {McClintock} J.~E., {Reid} M.~J., 2012, \apj, 745, L7

\bibitem[{Tchekhovskoy}, {Narayan}, \& {McKinney}(2011)]{tchekhovskoy11}
{Tchekhovskoy}, A., {Narayan} R., {McKinney} J.~C., 2011, \mnras, 418, 79

\bibitem[{Swank} et~al.(2010)]{swank10}
{Swank} J., {Kallman} T., {Jahoda} K., {Black} K., {Deines-Jones} P., {Kaaret} P., 2010, in {Bellazzini} R., {Costa} E., {Matt} G., {Tagliaferri} G., eds, {X-ray Polarimetry: A New Window in Astrophysics, Cambridge Univ. Press, Cambridge}, p.251

\bibitem[{Webster} \& {Murdin}(1972)]{webster72}
{Webster} L., {Murdin} P. 1972, Nat, 235, 37

\bibitem[{Zhang} et~al.(1997a)]{zhang97a}
{Zhang} S.~N., {Cui} W., {Chen} W., 1997a \apj, 482, L155

\bibitem[{Zhang} et~al.(1997b)]{zhang97b}
{Zhang} S.~N., {Cui} W., {Harmon} B.~A., {Paciesas} W.~S., {Remillard} R.~E., {van Paradijs} J., 1997b \apj, 477, L95



\end{thebibliography}


\appendix

\section{Luminosity Matching Model}\label{app:lum}

The functional form of the matching flux profile is given below (see \citealt{kulkarni11}, \citealt{pagethorne74} -- hereafter PT):
\small
\begin{align}\label{eq:fGPT}
f(r) = f_{\rm PT}(r)& - \frac{[(E^\dagger-\Omega L^\dagger)^2/\Omega_{,r}]_{r_{ie}}}{[(E^\dagger-\Omega L^\dagger)^2/\Omega_{,r}]_r}f_{\rm PT}(r_{ie}) \nonumber \\ 
& - \left[\frac{\Omega_{,r}}{(E^\dagger-\Omega L^\dagger)^2}\right]_r C,\qquad {\rm when}\; r>r_{ie} 
\end{align}
\normalsize
where $f(r) = 4 \pi r F_{\rm com}(r)/\dot{M}$ is the dimensionless flux, $L^\dagger$, $E^\dagger$, $\Omega$ are the specific angular momentum, energy-at-infinity, angular velocity (given by Eqs. 15f-h of PT), $f_{\rm PT}$ is the flux from the Page \& Thorne model (Eq. 15n of PT), and $C$ is a free parameter that is determined by the torque at the inner disc boundary.

The task now is to find an appropriate value of $C$ such that the GRMHD flux and Eq. \ref{eq:fGPT} match at $r_{ie}$.  For simplicity, we factor out the radial dependence from the last two terms of Eq. \ref{eq:fGPT} to get:
\begin{equation}\label{eq:fr}
f(r) = f_{\rm PT}(r) - \left[\frac{\Omega_{,r}}{(E^\dagger-\Omega L^\dagger)^2}\right]_{r} K, 
\end{equation}
where $K \equiv \left\{ [(E^\dagger-\Omega L^\dagger)^2/\Omega_{,r}]_{r_{ie}} f_{\rm PT}(r_{ie}) - C \right\}$ is just another constant.  To ensure continuity in the flux profile at $r_{ie}$, we require K to be:
\begin{equation}
K = \left[\frac{(E^\dagger-\Omega L^\dagger)^2}{\Omega_{,r}}\right]_{r_{ie}} \left[f_{\rm GRMHD}(r_{ie})-f_{\rm PT}(r_{ie})\right].
\end{equation}
Thus, our final matched flux profile is given by:
\begin{subnumcases}{\label{eq:Teff} \sigma_{\rm SB} T_{\rm eff}(r)^4 =}
  \frac{1}{-4\pi r u_t} \frac{dL_{\rm BL}(r)}{dr}, & $r\leq r_{ie}$, \label{eq:Teffin}\\
  \frac{\dot{M}}{4 \pi r} f(r), & $r > r_{ie}$, \label{eq:Teffout}
\end{subnumcases}
where $L_{\rm BL}(r)$ is the Boyer-Lindquist luminosity profile measured from the GRMHD simulations, and $f(r)$ is the dimensionless flux as shown in Eq. \ref{eq:fr}.


\section{Generalized Novikov \& Thorne Model}\label{app:vr}

To get the generalized \citetalias{NT} column density and radial velocity profile far out in the disc (to extend the GRMHD simulated disc beyond $r_{ie}$), we solve the following set of vertical structure equations (with a one-zone model for the vertical structure):
\begin{enumerate}
\begin{subequations} \label{eq:SigEqns}
\item Vertical Pressure Balance:
\begin{equation}
\frac{dP_{\rm tot}}{dz} = \rho g_\perp, \nonumber
\end{equation}
has the vertically integrated form of:
\begin{equation}
\frac{P_{\rm tot}}{h} = \rho Q h, \label{eq:pressurebal}
\end{equation}
where h is the vertical scale height, and $Q=g_\perp/z$ is the prescription for the local vertical gravity (as defined in Eq. \ref{eq:Qdef}).\\

\item Radiative Transfer:

From the second moment of radiative transfer equation (see \S1.8 of \citealt{rybicki79}):
\begin{equation}
\frac{dP_{\rm rad}}{d\tau}=\frac{F_{\rm rad}}{c}, \nonumber
\end{equation}
we transform the $d\tau$ optical depth differential to a column mass differential $dm$ via $d\tau = \kappa \cdot dm$, where $\kappa$ is the Rosseland mean opacity. We integrate the column mass from the surface to the disc midplane (from $m=0$ to $m=m_{\rm tot}=\Sigma/2$). We also assume that the flux profile linearly increases with mass away from the midplane (i.e. $F(m) = F_{\rm tot}\left(1-m/m_{\rm tot}\right)$ -- this linearity assumption is also used in \textsc{tlusty}).  The vertically integrated radiative diffusion equation thus becomes
\begin{equation}\label{eq:radiativeDiffusion}
P_{\rm rad} = \frac{aT^4}{3}=\frac{F_{\rm tot} \kappa \Sigma}{4c},
\end{equation}
where $a$ is the radiation constant, and $T$ is radiation temperature.  Note: The factor 2 discrepancy in Eq. \ref{eq:radiativeDiffusion} compared to the standard prescription for radiative diffusion stems from the linearity assumption in $F(m)$.\\

\item Stress:

We adopt an $\alpha$ prescription for the stress where $t_{\hat{\phi}\hat{r}}=\alpha p_{\rm tot}$.  Vertical integration yields
\begin{equation}
W \equiv \int_{-h}^{h} t_{\hat{\phi}\hat{r}} dz = 2 h \alpha P_{\rm tot}.
\end{equation}
\\

\item Viscous Heating:

Through the energy equation for viscous heating, it is possible to link the heating flux with the vertically integrated stress (c.f. 5.6.7-12 of \citealt{NT}).  The resulting expression is
\begin{equation}
F_{\rm tot}=\frac{3}{4} \Omega_k R_F(r) W,
\end{equation}
with the dimensionless relativistic factor $R_F(r)$ defined as
\begin{equation}
R_F(r) = \frac{1-2/r_*+a_*^2/r_*^2}{1-3/r_*+2a_*/r_*^{3/2}} \nonumber.
\end{equation}
\\

\item Equation of State:

We ignore magnetic pressure in this analysis, and adopt an ideal gas equation of state, yielding
\begin{equation}
P_{\rm tot}=\frac{\rho k_B T}{\mu} + \frac{aT^4}{3},
\end{equation}
where $\mu$ is the mean particle weight of the fluid.
\\

\item Column Density

In this one-zone model, the column density represents the quantity
\begin{equation}
\Sigma = \int_{-h}^{h}\rho dz = 2h\rho.
\end{equation}
\end{subequations}

\end{enumerate}

Our goal is to solve Eqs. \ref{eq:SigEqns} with unknowns ($P_{\rm tot}, \rho, T, \Sigma, h$), given the values of ($\kappa, \mu, \alpha, F_{\rm tot}, Q, M, r$).  After some algebra, we obtain an expression for $\Sigma$ that solves Eqs. \ref{eq:SigEqns}.  We find $\Sigma$ by solving for the real root of the following polynomial in $x$:
\begin{align} \label{eq:sigpoly}
(F_1)^2 & -\left[\frac{F_1}{4F_3}\right] x^4 - \left[2\left(\frac{3}{8}\right)^{1/4}F_1 F_2\right] x^5 \nonumber \\
& + \left[\left(\frac{3}{8}\right)^{1/2}(F_2)^2\right] x^{10} = 0,
\end{align}
where $x = \Sigma^{1/4}$ and the polynomial coefficients $F_1,F_2,F_3$ depend only on the given values:
\begin{subequations}

\begin{equation}
F_1 = \frac{W}{\alpha} = \frac{4 F_{\rm tot}}{3\alpha R_F \Omega_k},
\end{equation}

\begin{equation}
F_2 = \left(\frac{\kappa F_{\rm tot}}{2 \sigma_{\rm SB}}\right)^{1/4} \left(\frac{k_B}{\mu}\right),
\end{equation}

\begin{equation}
F_3 = \frac{c^2 Q}{\kappa^2 F_{\rm tot}^2}.
\end{equation}

\end{subequations}

We then solve Eq. \ref{eq:sigpoly} at different radii to obtain $\Sigma(r)$ for ($r > r_{ie}$).  We pick $\kappa=\kappa_{es}=0.3383$\,cm$^2$\,g$^{-1}$ and $\mu = 0.615\,m_{H}$, which corresponds to a fluid composition of 70 per cent Hydrogen, 30 per cent Helium by mass.  Given $\Sigma(r)$, we get the mass averaged radial velocity by solving for $\tilde{u}^r$ in the mass conservation equation $\dot{M}=2\pi r \Sigma \tilde{u}^r$.


\section{Interpolation Methods}\label{app:interpolation}

All interpolated quantities are computed by linear interpolation in log space (i.e. for two variables $x$, and $y$, we interpolate linearly on $\log(x)$ and $\log(y)$).  We have chosen this interpolation scheme since it can perfectly capture all power-law scalings.  For quantities obtained by interpolating on \textsc{tlusty} annuli (such as the annuli vertical structure), we employ trilinear interpolation over the 3 annuli parameters $\log(T_{\rm eff}),\log(\Sigma)$, and $\log(Q)$ (i.e. each parameter is linearly interpolated seperately in log space).

The interpolation of spectra $I_{\nu}$ is handled by a more complicated method to account for the shape of the Planck function.  For an optically thick medium with scattering, the resultant spectrum takes on the form of a modified blackbody (c.f. Eq. \ref{eq:modBB}).  Since the spectra doesn't depend very sensitively on $\Sigma$, or $Q$, we apply the log-linear interpolation discussed above for these two annuli parameters. For $T_{\rm eff}$, we use a more complicated interpolation scheme.  Rather than follow \citet{davis05} (they applied a linear interpolation on the brightness temperature, computed with a fixed spectral hardening factor of $f=2.0$), we switch between three different interpolation methods that each work for all choices of $f$.  The three methods are applied to different frequency ranges of the spectrum.

\begin{itemize}
\item Method 1: At low frequencies, we have that $I_v \propto T_{\rm eff}$ in the Raleigh-Jeans tail, and thus we can apply the linear interpolation method.  
\item Method 2: At high frequencies, we apply linear interpolation on $1/T_{\rm eff}$ and $\log(I_\nu)$.  The motivation is that for a blackbody-like spectrum in the high-frequency Wien limit, the specific intensity as a function of temperature scales as $\log(I_v) \propto -1/T_{\rm eff}$.  
\item Method 3: For intermediate frequencies, we take a non-linear combination of the two interpolation results so that the transition between interpolation methods is smooth.
\end{itemize}
\begin{figure}
\begin{center}  
\includegraphics[width=0.5\textwidth]{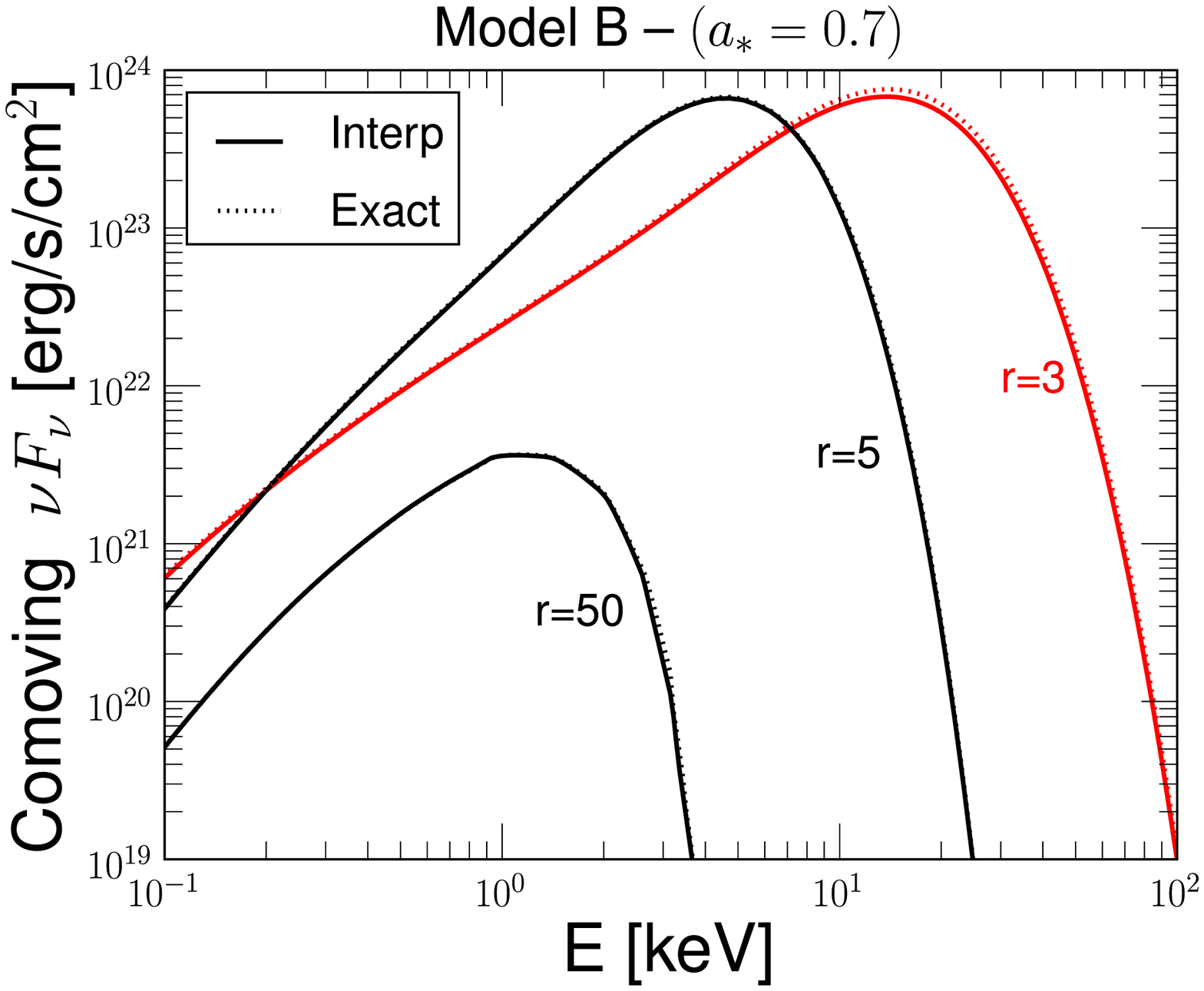} 
\caption{Comparison of annuli spectra obtained through the interpolation method (solid) and an exact calculation (dashed) for the spin $a_*=0.7$ disc.  The black spectra represent annuli outside the plunging region, whereas the red spectrum corresponds to a plunging region annulus.\label{fig:spectraInterpolate}}
\end{center}
\end{figure}
Denoting the interpolated intensities using methods 1 and 2 as $I_{1}(\nu)$ and $I_{2}(\nu)$ respectively, the expression for the final combined interpolation method is:
\begin{subnumcases}{\label{eq:Inu} I_\nu(\nu)=}
  I_{1}(\nu), & $\nu \leq \nu_1$, \nonumber  \\
  \exp\left\{ \frac{\ln(\nu_2)-\ln(\nu)}{\ln(\nu_2)-\ln(\nu_1)}  \ln\left[I_{1}\right(\nu)] \right. & \nonumber \\
  \quad + \left. \frac{\ln(\nu)-\ln(\nu_1)}{\ln(\nu_2)-\ln(\nu_1)} \ln\left[I_{2}(\nu)\right]\right\}, & $\nu_1 < \nu < \nu_2$, \nonumber \\
  I_{2}(\nu), & $\nu \geq \nu_2$. \nonumber
\end{subnumcases}

We have set $\nu_1 = 0.07\,\nu_{\rm max}$, and $\nu_2 = 2.3\,\nu_{\rm max}$, where $\nu_{\rm max}$ is the frequency that maximizes $I_\nu$ in the spectra of the lowest $T_{\rm eff}$ annuli in the bracketing set used for the interpolation.  These frequency boundary values were determined by minimizing the interpolation error for a pure Planck function.  For a perfect modified Planck function with constant $f$, the chosen $\nu_1$ and $\nu_2$ correspond to a maximum interpolation error of 0.08 per cent in $I_\nu$ for our grid with temperature spacings $\Delta \log_{\rm 10}(T_{\rm eff})=0.1$.  In practice, the interpolation error after interpolating over the 3 annuli parameters and emission angle is $\sim$1 per cent (see Fig. \ref{fig:spectraInterpolate}), except for annuli in the plunging region, where the flux error grows to $\sim$10 per cent (the assumption of a constant $f$ is violated for annuli in the plunging region).  The main advantage of this interpolation method is it provides interpolated spectra that peak very close to the correct frequencies, and the method works equally well across all spectral hardening factors.\\


\section{An Alternative GRMHD Luminosity Profile}\label{app:cooling}
We wish to generate a disc luminosity profile that has the following desirable properties: 1) it is based on the GRMHD dissipation profile rather than cooling profile, and 2) it produces a disc whose entropy profile is self-consistent with the \textsc{TLUSTY} annuli.   We accomplish this task by considering the physics of energy advection.  The idea is to solve for the radiative cooling rate, taking into account both the dissipative heating rate (measured from the GRMHD simulations) and the heat advection rate (computed using the vertical structure from \textsc{TLUSTY} annuli).  The energy balance equation relating these three quantities is:
\begin{equation}\label{eq:advection}
\hat{q}_{{\rm adv}} =\hat{q}_{{\rm heat}}-\hat{q}_{{\rm cool}},
\end{equation}
where $\hat{q}_{{\rm adv}}$ is the advected heating rate per unit volume, $\hat{q}_{{\rm heat}}$ the viscous heating rate per unit volume, and $\hat{q}_{{\rm cool}}$ is the cooling rate per unit volume (assumed to be purely radiative -- i.e. we ignore conduction and convection).  We take the convention that all hatted quantities are measured in the comoving frame of the fluid.  The advective heating rate is obtained by evaluating $\hat{q}_{{\rm adv}} = \hat{\rho} \hat{T} (d\hat{s}/d\tau)$, where $\hat{\rho}$ is the rest mass density, $\hat{T}$ is the gas temperature, $\hat{s}$ is the specific gas entropy of the fluid, and $\tau$ is the fluid's proper time.  Making the approximation that we have stationary axisymmetric flow that is devoid of motion perpendicular to the midplane, we have $d\hat{s}/d\tau = u^r d\hat{s}/dr$  via the chain-rule ($d\hat{s}/dr$ and $u^r$ are evaluated in BL-coordinates).  Vertically integrating Eq. \ref{eq:advection} while making use of $\hat{q}_{{\rm adv}} = \hat{\rho} \hat{T} u^{r} (d\hat{s}/dr)$ yields:
\begin{align}
\hat{Q}_{{\rm adv}} & = \hat{Q}_{{\rm heat}}^{\rm sim}-\hat{Q}_{{\rm cool}}\label{eq:verticaladvection} \\ 
& = \int_{z=-\infty}^{+\infty} \hat{\rho} \hat{T} u^r \left(\frac{d\hat{s}}{dr}\right) dz\label{eq:verticaladvection2}, 
\end{align}
where $\hat{Q}_{{\rm adv}}$ is the vertically integrated heat advection rate (evaluated with TLUSTY vertical structure), $\hat{Q}_{{\rm heat}}^{\rm sim}$ is the vertically integrated viscous heating rate (measured from the GRMHD simulated discs), and $\hat{Q}_{{\rm cool}}$ is the net radiative cooling flux that escapes.  For simplicity in Eq. \ref{eq:verticaladvection2}, we adopt the following constant mass-averaged radial velocity:
\begin{equation}
u^r=\frac{\int_{\theta=0}^{\pi} \hat{\rho} u^r_{\rm sim} \sqrt{-g} d\theta}{\int_{\theta=0}^{\pi} \hat{\rho}\sqrt{-g} d\theta}. 
\end{equation}
where $u^r_{\rm sim}$ represents the pointwise radial velocities measured from GRMHD simulations.\\

\emph{Our goal is to find the value of $\hat{Q}_{{\rm cool}}$ that satisfies Eq. \ref{eq:verticaladvection}}, given that we can measure $\hat{Q}_{{\rm heat}}^{\rm sim}$ from the simulations and calculate $\hat{Q}_{\rm adv}$ from Eq. \ref{eq:verticaladvection2}.

\subsection{Obtaining \texorpdfstring{$\hat{Q}_{\rm heat}^{\rm sim}$}{Qheatsim} (the GRMHD dissipation profile) }\label{app:dissipation}

Unfortunately, the GRMHD simulations that we ran were not set up to directly keep track of the numerical dissipation $\hat{Q}_{{\rm heat}}^{\rm sim}(r)$.  Despite this lack of information, we can still estimate the dissipation indirectly by running the argument in Eq. \ref{eq:verticaladvection} backwards; we solve for the vertically integrated dissipative heating rate
\begin{equation}\label{eq:dissipation}
\hat{Q}_{{\rm heat}}^{\rm sim} = \hat{Q}_{{\rm adv}}^{\rm sim}+\hat{Q}_{{\rm cool}}^{\rm sim}.
\end{equation}
The `sim' superscript is used to denote quantities derived solely from the GRMHD simulations (i.e. these quantities are independent of TLUSTY).  Analogous to Eq. \ref{eq:verticaladvection2}, the vertically integrated\footnote{To maintain consistency with the simulation cooling flux $\hat{Q}_{{\rm cool}}^{\rm sim}$ (as calculated by Eqs. \ref{eq:lum} and \ref{eq:flux}, which only considers bound disc fluid), the integral in Eq. \ref{eq:verticaladvectionBL} is likewise only taken over the bound fluid.} GRMHD advective heating rate is obtained by
\begin{equation}\label{eq:verticaladvectionBL}
\hat{Q}_{{\rm adv}}^{\rm sim} = \int_{\theta=0}^{\pi} \hat{\rho} \hat{T} \left(\frac{d\hat{s}}{dr}\right) u^r \sqrt{-g} d\theta.
\end{equation}
For simplicity, we first apply azimuthal and time averaging to all simulation quantities used in Eq. \ref{eq:verticaladvectionBL}.  Since the GRMHD simulations are dimensionless ($k/\mu=1$) and employ an ideal gas equation of state, we have that $\hat{T}=\hat{P}_{\rm gas}/\hat{\rho}$, and $\hat{s}=(\Gamma-1)^{-1}\ln(\hat{P}_{\rm gas}/\hat{\rho}^\Gamma)$.  The GRMHD cooling flux is simply $\hat{Q}_{{\rm cool}}^{\rm sim} = F_{\rm com}$ where $F_{\rm com}$ is the comoving disc flux, as given by Eq. \ref{eq:flux}.

\subsection{Net result of the luminosity calculation}

The final goal is to solve for the value of $\hat{Q}_{\rm cool}$ that satisfies Eq. \ref{eq:verticaladvection}.  The physical interpretation of this newly derived $\hat{Q}_{\rm cool}$ is simply the cooling rate corresponding to a disc with the vertical structure given by \textsc{TLUSTY} that is heated up according to the GRMHD dissipation profile.  Putting everything together, substituting Eq. \ref{eq:dissipation} for $\hat{Q}_{\rm heat}^{\rm sim}$ into Eq. \ref{eq:verticaladvection} gives $\hat{Q}_{\rm cool}$ as:
\begin{equation}\label{eq:netDissResult}
\hat{Q}_{\rm cool} = \hat{Q}_{{\rm cool}}^{\rm sim}+\hat{Q}_{{\rm adv}}^{\rm sim}-\hat{Q}_{{\rm adv}}.
\end{equation}
$\hat{Q}_{{\rm adv}}^{\rm sim}$ and $\hat{Q}_{{\rm cool}}^{\rm sim}$ are computed from Eq. \ref{eq:verticaladvectionBL} and Eq. \ref{eq:flux} respectively. $\hat{Q}_{{\rm adv}}$ uses the \textsc{TLUSTY} vertical structure, and is evaluated via Eq. \ref{eq:verticaladvection2}.  However, the process of locating the correct value for $\hat{Q}_{{\rm cool}}$ that solves Eq. \ref{eq:netDissResult} is complicated by two factors:

\begin{enumerate}
\item $\hat{Q}_{{\rm adv}}$ on the right hand side is a function of $\hat{Q}_{{\rm cool}}$. $\hat{Q}_{{\rm adv}}$ indirectly depends on $\hat{Q}_{{\rm cool}}$ through the annuli vertical structure (since $\hat{T}_{{\rm eff}}=[\hat{Q}_{{\rm cool}}/\sigma_{{\rm SB}}]^{1/4}$ is an annuli parameter).  The strategy that we employ to find the correct $\hat{Q}_{{\rm cool}}$ is a bisection method; we start with the two initial guesses for $\hat{Q}_{{\rm cool}}$ that bracket the relation in Eq. \ref{eq:netDissResult} (which we have empirically found to be monotonic with $\hat{Q}_{{\rm cool}}$).  We then bisect on this $\hat{Q}_{{\rm cool}}$ interval until Eq. \ref{eq:netDissResult} is satisfied.\\

\item To compute the $d\hat{s}/dr$ term, we need to know $\hat{Q}_{{\rm cool}}$ (or equivalently $\hat{T}_{\rm eff}$) for two neighboring annuli.  However the process outlined above (in point 1) only lets us solve $\hat{Q}_{{\rm cool}}$ for a single annulus.  The resolution to this problem is to choose a value of $\hat{Q}_{{\rm cool}}$ for the outermost annulus as a boundary condition.  If there are a total of $N$ annuli, then given $\hat{Q}_{{\rm cool}}$ for the $N^{\rm th}$ annuli, we can solve $\hat{Q}_{{\rm cool}}$ for the $(N-1)^{\rm th}$ annuli.  This process can be iterated to obtain $\hat{Q}_{{\rm cool}}$ for all remaining interior annuli.  We set $\hat{Q}_{{\rm cool}}=\hat{Q}_{\rm heat}$ as the boundary condition for our outermost annuli since far into the disc, energy advection becomes negligible (in other words, $\hat{Q}_{{\rm adv}}\rightarrow 0$ far out in the disc, which implies $\hat{Q}_{{\rm cool}}\rightarrow \hat{Q}_{\rm heat}$ from Eq. \ref{eq:verticaladvection}).\\
\end{enumerate}

Running through the above two steps allows us to solve for $\hat{Q}_{{\rm cool}}$, and hence obtain a second measure of the disc luminosity (labelled as GRMHD2 in all plots and tables).  Note that the GRMHD2 model is more self-consistent with TLUSTY in Fig. \ref{fig:s}.  The TLUSTY2 model (computed from the newly derived GRMHD2 disc luminosities) and the intrinsic GRMHD entropy profiles agree fairly well when $r>r_{\rm ISCO}$.

\end{document}